\documentclass[final]{elsarticle}

\usepackage{caption}
\usepackage{adjustbox}
\usepackage{float}
\usepackage{lineno,hyperref}
\modulolinenumbers[5]

\journal{PLoS Computational Biology}

\makeatletter
\def\@author#1{\g@addto@macro\elsauthors{\normalsize%
    \def\baselinestretch{1}%
    \upshape\authorsep#1\unskip\textsuperscript{%
      \ifx\@fnmark\@empty\else\unskip\sep\@fnmark\let\sep=,\fi
      \ifx\@corref\@empty\else\unskip\sep\@corref\let\sep=,\fi
      }%
    \def\authorsep{\unskip,\space}%
    \global\let\@fnmark\@empty
    \global\let\@corref\@empty  
    \global\let\sep\@empty}%
}

\usepackage{amsmath,amsfonts}

\newcommand{\argmax}{\arg \max}

\begin{document}

\begin{frontmatter}

\title{Breakdown of local information processing may underlie isoflurane anesthesia effects}


\author{Patricia Wollstadt\corref{mycorrespondingauthor}\fnref{bic}}
\author{Kristin K. Sellers\fnref{unc_psych,unc_neurobiol}}
\author{Lucas Rudelt\fnref{mpi}}
\author{Viola Priesemann\corref{mycorrespondingauthor}\fnref{mpi,bccn}}
\author{Axel Hutt\fnref{dwd,read}}
\author{Flavio Fr{\"o}hlich\fnref{unc_psych,unc_neurobiol,unc_cellbiol,unc_biomedeng,unc_neurosci,unc_neurol}}
\author{Michael Wibral\fnref{bic}}

\fntext[bic]{MEG Unit, Brain Imaging Center, Goethe University, Frankfurt/Main, Germany}
\fntext[unc_psych]{Department of Psychiatry, University of North Carolina at Chapel Hill, Chapel Hill NC 27599, USA}
\fntext[unc_neurobiol]{Neurobiology Curriculum, University of North Carolina at Chapel Hill Chapel Hill NC 27599, USA}
\fntext[mpi]{Max Planck Institute for Dynamics and Self-Organization, G\"ottingen, Germany}
\fntext[bccn]{Bernstein Center for Computational Neuroscience, BCCN, G\"ottingen, Germany}
\fntext[dwd]{Deutscher Wetterdienst, Section FE 12 - Data Assimilation, Offenbach/Main, Germany}
\fntext[read]{Department of Mathematics and Statistics, University of Reading, Reading, United Kingdom}
\fntext[unc_cellbiol]{Department of Cell Biology and Physiology, University of North Carolina at Chapel Hill, Chapel Hill NC 27599, USA}
\fntext[unc_biomedeng]{Department of Biomedical Engineering, University of North Carolina at Chapel Hill, Chapel Hill NC 27599, USA}
\fntext[unc_neurosci]{Neuroscience Center, University of North Carolina at Chapel Hill, Chapel Hill NC 27599, USA}
\fntext[unc_neurol]{Department of Neurology, University of North Carolina at Chapel Hill, Chapel Hill NC 27599, USA}

\cortext[mycorrespondingauthor]{Corresponding authors}
\ead{p.wollstadt@stud.uni-frankfurt.de, viola@nld.ds.mpg.de}

\begin{abstract}
The disruption of coupling between brain areas has been suggested as the mechanism underlying loss of consciousness in anesthesia. This hypothesis has been tested previously by measuring the information transfer between brain areas, and by taking reduced information transfer as a proxy for decoupling. Yet, information transfer is a function of the amount of information available in the information source---such that transfer decreases even for unchanged coupling when less source information is available. Therefore, we reconsidered past interpretations of reduced information transfer as a sign of decoupling, and asked whether impaired local information processing leads to a loss of information transfer. An important prediction of this alternative hypothesis is that changes in locally available information (signal entropy) should be at least as pronounced as changes in information transfer. We tested this prediction by recording local field potentials in two ferrets after administration of isoflurane in concentrations of 0.0~\%, 0.5~\%, and 1.0~\%.

We found strong decreases in the source entropy under isoflurane in area V1 and the prefrontal cortex (PFC)---as predicted by our alternative hypothesis. The decrease in source entropy was stronger in PFC compared to V1. Information transfer between V1 and PFC was reduced bidirectionally, but with a stronger decrease from PFC to V1. This links the stronger decrease in information transfer to the stronger decrease in source entropy---suggesting reduced source entropy reduces information transfer. This conclusion fits the observation that the synaptic targets of isoflurane are located in local cortical circuits rather than on the synapses formed by interareal axonal projections. Thus, changes in information transfer under isoflurane seem to be a consequence of changes in local processing more than of decoupling between brain areas. We suggest that source entropy changes must be considered whenever interpreting changes in information transfer as decoupling.
\end{abstract}

\end{frontmatter}

\section*{Author Summary}   
Currently we do not understand how anesthesia leads to loss of consciousness (LOC). One popular idea is that we loose consciousness when brain areas lose their ability to communicate with each other – as anesthetics might interrupt transmission on nerve fibers coupling them. This idea has been tested by measuring the amount of information transferred between brain areas, and taking this transfer to reflect the coupling itself. Yet, information that isn't available in the source area can’t be transferred to a target. Hence, the decreases in information transfer could be related to less information being available in the source, rather than to a decoupling. We tested this possibility measuring the information available in source brain areas and found that it decreased under isoflurane anesthesia. In addition, a stronger decrease in source information lead to a stronger decrease of the information transfered. Thus, the input to the connection between brain areas determined the communicated information, not the strength of the coupling (which would result in a stronger decrease in the target). We suggest that interrupted information processing within brain areas has an important contribution to LOC, and should be focused on more in attempts to understand loss of consciousness under anesthesia.


\section*{Introduction}

To this day it is an open question in anesthesia research how general anesthesia leads to loss of consciousness (LOC). Several recent theories agree in proposing that anesthesia-induced LOC may be caused by the disruption of long range inter-areal information transfer in cortex \cite{dehaene2011GNW,tononi2004iit,imas2005,hudetz2006,alkire2008}---a hypothesis supported by a series of recent studies  \cite{imas2005,ku2011STE,lee2013,jordan2013,untergehrer2014}. In all of these studies, information transfer is quantified using transfer entropy \cite{schreiber2000}, an information theoretic measure, which has become a quasi-standard for the estimation of information transfer in anesthesia research, or by transfer entropy's linear implementation as a Granger causality. In many of these studies reduced information transfer has been interpreted as a sign of inter-areal long range connectivity being disrupted by anesthesia.

Yet, information transfer between a source of information and a target depends on information (entropy) being available at the source in the first place. Considering constraints of this kind we can easily conceive of cases where a decrease in information transfer under anesthesia is observed despite unchanged long range coupling, e.g., when the available information at the source decreases due to an anesthesia-related change in \textit{local} information processing. Ultimately, this dissociation between information transfer and causal coupling just reflects that information transfer is \textit{one} possible consequence of physical coupling, but not identical to it \cite{Ay2008,lizier2010,chicharro2012}.

Therefore, we consider it necessary to evaluate the hypothesis that the reduced inter-areal information transfer observed under anesthesia possibly originates from disrupted information processing in local circuits rather than from disrupted long range connectivity. This alternative hypothesis receives additional support for the case of isoflurane, which potentiates agonist actions at $GABA_A$-receptors and inhibits nicotinic acetylcholine (nAChR) receptors. Conversely, evidence on direct inhibitory effects of isoflurane on AMPA and NMDA synapses, which are the dominant mediators of long-range cortico-cortical interactions, is sparse at best (see table 2 in \cite{krasowski1999}). Under the alternative hypothesis of changed local information processing, a decrease in transfer entropy under anesthesia must be accompanied by:

\begin{enumerate}
	\item a reduction in locally available information per brain area, i.e. in the sources of information transfer,
	\item and the fact that the strongest decrease in locally available information is found at the source of the link with the strongest decrease in information transfer, rather than at its target (i.e. the end point),
\end{enumerate}

Here, we perform tests of these predictions by estimating local information processing in and information transfer between local field potentials (LFPs) simultaneously recorded from primary visual cortex (V1) and prefrontal cortex (PFC) of two ferrets under different levels of isoflurane. We quantify local information processing by estimating the signal entropy (measuring available information) and quantified information transfer between recording sites by estimating transfer entropy. Additionally, to demonstrate the effect of reduced source entropy on transfer entropy, we estimated transfer entropy on simulated data with a constant coupling between processes, but a varying source entropy.

To better understand potential changes in local information processing we also quantified the active information storage, a measure of the information available at a recording site that can be predicted from past signals at that site, i.e. the information stored from past to present.

Because the estimation of such information theoretic quantities from finite data is difficult in general, we employ two complementary strategies: (i) probability density estimation based on nearest-neighbor searches in continuous data, and (ii) Bayesian estimation based on discretized data.

We also test whether the previously reported decrease of transfer entropy under anesthesia can indeed be replicated when avoiding some recently identified pitfalls in estimation of information transfer related to the use of symbolic time series, suboptimal embedding of the time series, and the use of net transfer entropy without identification of the individual information transfer delays  (for problems related to these approaches see \cite{wibral2013timing}).

Our results provide first evidence for the alternative hypothesis of altered local information processing causing reduced information transfer, as the above predictions were indeed met. We suggest to consider the alternative hypothesis as a serious candidate mechanism for LOC, and to use causal interventions to gather further experimental evidence.

Preliminary results for this study were published in abstract form in \cite{wollstadt2015IEEE}.

\section*{Results}
\subsection*{Existence of information transfer between recording sites}
Before analyzing differences in information transfer induced by isoflurane, we tested for the existence of significant information transfer between the recording sites (table \ref{tab:te_sign}). For both animals, $TE_{SPO}$ was significant in the top-down direction, while the bottom-up direction was significant for animal 1 only. A non-significant information transfer in bottom-up direction for animal 2 may be explained by the dark experimental environment, i.e., the lack of visual input.

\begin{table}[!ht]
	\caption{Results significance test of $TE_{SPO}$ estimates in both animals and for both directions of interaction.}
	\label{tab:te_sign}
	\begin{tabular}{llc}
		\hline
		\textbf{animal}& \textbf{direction of interaction} & \textbf{$p$-value} \\ \hline
		Ferret 1 & PFC $\rightarrow$ V1 & \textless0.01$^{**}$ \\
		         & V1 $\rightarrow$ PFC & \textless0.05$^{*}$ \\
		Ferret 2 & PFC $\rightarrow$ V1 & \textless0.05$^{*}$ \\
		         & V1 $\rightarrow$ PFC & 0.2262 n.s. \\ \hline
		\multicolumn{2}{l}{$^* p<0.05$; $^{**} p<0.01$; $^{***} p<0.001$; Bonferroni-corrected}
	\end{tabular}
	\vspace*{-4pt}
\end{table}

%

\subsection*{Changes in information theoretic measures under anesthesia}
Overall, for higher isoflurane levels both locally available information and information transfer were decreased, while information storage in local activity increased.

As the estimation of information theoretic measures from finite length neural recordings poses a considerable challenge we present detailed, converging results from two complementary strategies to deal with this challenge---nearest-neighbor based estimators, and a Bayesian approach to entropy estimation suggested by Nemenman, Shafe, and Bialek (NSB-estimator) \cite{nsb,nsb2}. This latter approach required a discretization of the continuous-valued LFP data, but yields principled control of bias, while the first approach allows the estimation of information-theoretic measures directly from continuous data, and thus conserves the information originally present in those data. Statistical testing was performed using a nonparametric permutation ANOVA (pANOVA), and a linear mixed model (LMM). The LMM approach was used in addition to the main pANOVA for the purpose of comparison to older studies using parametric statistics.

\begin{figure}[!h]
    \includegraphics{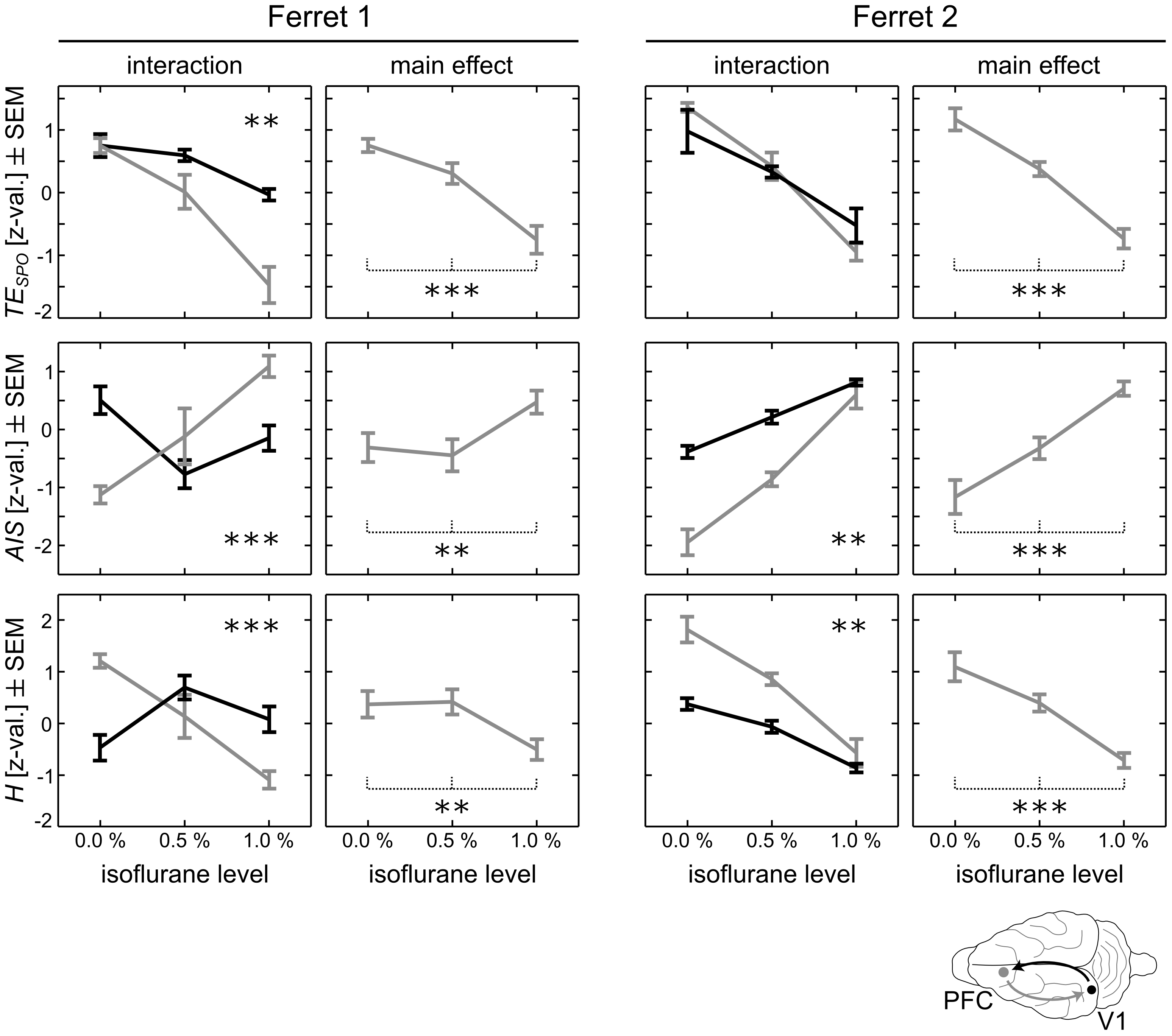}
\caption{{\bf pANOVA results for nearest-neighbor based estimates of transfer entropy ($TE_{SPO}$), active information storage ($AIS$), and entropy ($H$).}
Left columns show interactions \textit{isoflurane level x direction} and \textit{isoflurane level x recording site} for both animals; right columns show main effects \textit{isoflurane level}. Grey lines in interaction plots indicate $TE_{SPO}$ from prefrontal cortex (PFC) to primary visual areas (V1), or $H$ and $AIS$ in PFC; black lines indicate $TE_{SPO}$ from V1 to PFC, or $H$ and $AIS$ in V1. Error bars indicate the standard error of the mean (SEM); stars indicate significant interactions or main effects ($^* p<0.05$; $^{**} p<0.01$; $^{***} p<0.001$). Axis units for all information theoretic measures based on continuous variables are z-normalized values across conditions.}
\label{fig:pANOVA_main_effects_all}
\end{figure}

\paragraph*{Results based on next neighbor-based estimation from continuous data.} For higher isoflurane levels, we found an overall reduction in the locally available information  ($H$), and in the information transfer ($TE_{SPO}$). We found an increase in the locally predictable (stored) information ($AIS$) (table \ref{tab:pANOVA} and Fig. \ref{fig:pANOVA_main_effects_all}).

\begin{table}[!ht]
\caption{Results of permutation analysis of variance for information theoretic measures ($p$-values).}
\label{tab:pANOVA}
\begin{tabular}{llcc}
\hline
\textbf{measure} & \textbf{effect} & \textbf{Ferret 1} & \textbf{Ferret 2} \\ \hline

$TE_{SPO}$  & \textit{isoflurane level} & \textless0.0001$^{***}$ & \textless0.0001$^{***}$ \\
    & \textit{direction} & 0.0003$^{***}$ &  0.9345 \\
    & \textit{interaction} & 0.0052$^{**}$ & 0.2486$^{a}$ \\ \hline

$AIS$ & \textit{isoflurane level} &  0.0017$^{**}$ & \textless0.0001$^{***}$ \\
    & \textit{recording site} & 0.6774 &  \textless0.0001$^{***}$ \\
    & \textit{interaction} & \textless0.0001$^{***}$  & 0.0029$^{**}$ \\ \hline

$H$   & \textit{isoflurane level} &  0.0010$^{**}$ & \textless0.0001$^{***}$ \\
    & \textit{recording site} & 0.9326 &  \textless0.0001$^{***}$ \\
    & \textit{interaction} & \textless0.0001$^{***}$  & 0.0243$^{*}$ \\ \hline
	\multicolumn{4}{l}{$^* p<0.05$; $^{**} p<0.01$; $^{***} p<0.001$;} \\
	\multicolumn{4}{l}{$^a$This effect was significant when using LMM for statistical testing} \\
\end{tabular}
\vspace*{-4pt}
\end{table}

In general, $H$ decreased in both animals under higher isoflurane levels (main effect \textit{isoflurane level}, $p<0.01^{**}$ for ferret 1 and $p<0.001^{***}$ for ferret 2), indicating a reduction of locally available information for higher isoflurane concentrations. Yet, an isoflurane concentration of 0.5~\% (abbreviated as \textit{iso 0.5~\%}, with other concentrations abbreviated accordingly) led to a slight increase in $H$ for ferret 1, followed by a decrease for concentration \textit{iso 1.0~\%}, which was below initial entropy values. This rise in $H$ in condition \textit{iso 0.5~\%} was present only in V1 of ferret 1, while $H$ decreased monotonically in PFC. In ferret 2, $H$ increased monotonically in both recording sites, with a stronger decrease in PFC. The interaction effect (\textit{isoflurane level x brain region}) was significant for both animals ($p<0.001^{***}$ for ferret 1 and $p<0.01^{**}$ for ferret 2).

The information transfer as measured by the self prediction optimal transfer entropy ($TE_{SPO}$, \cite{wibral2013timing}) decreased significantly with higher levels of isoflurane in both animals (main effect \textit{isoflurane level}, $p<0.001^{***}$), indicating an overall reduction in information transfer. This reduction was stronger in the top-down direction $PFC \rightarrow V1$ (significant interaction effect \textit{isoflurane level x direction} in ferret 1, $p<0.01^{**}$). In ferret 2 this interaction was not significant in the permutation ANOVA (pANOVA) on aggregated data, but was highly significant using the LMM approach (see next section).

The stored information as measured by the active information storage ($AIS$, \cite{lizier2012LAIS}) increased in both animals under higher isoflurane levels (main effect \textit{isoflurane level}, $p<0.01^{**}$ for ferret 1 and $p<0.001^{***}$ for ferret 2), indicating more predictable information in LFP signals under higher levels of isoflurane. In ferret 1 the concentration \textit{iso 0.5~\%} led to a slight decrease in $AIS$, followed by an increase compared to initial levels for concentration \textit{iso 1.0~\%}. This initial decrease in $AIS$ in condition \textit{iso 0.5~\%} was present only in V1 of this animal, while in its PFC $AIS$ increased monotonically. In ferret 2, $AIS$ increased monotonically in both recording sites, with a stronger increase in PFC. The interaction effect was significant for both animals ($p<0.001^{***}$ for ferret 1 and $p<0.01^{**}$ for ferret 2). Overall, $AIS$ behaved complementary to $H$ for all animals and isoflurane levels, despite the fact that AIS is one component of $H$ \cite{lizier2012LAIS}.

\paragraph*{Alternative statistical testing using linear mixed models} We additionally performed a parametric test, using linear mixed models (LMM) on non-aggregated data from individual epochs of recording sessions (adding \textit{recording} as additional random factor to encode the recording session) to enable a comparison to results from earlier studies using parametric testing.

For both animals, model comparison showed a significant main effect of factor \textit{isoflurane level} on $TE_{SPO}$, $AIS$, and $H$ as well as a significant interaction of factors \textit{isoflurane level} and \textit{direction} (see table \ref{tab:lmm_modcomp}). The only exception to this was the factor direction in the evaluation of $TE_{SPO}$ in Ferret 2. Thus, the results of this alternative statistical analysis were in agreement with those from the pANOVA. The  detailed table \ref{tab:lmm_modcomp} reports the Bayesian Information Criterion with $BIC =  -2 \log p(\mathbf{x}| M, \hat{a}) + k(a) log(N)$, where $\mathbf{x}$ are the observed realizations of the data, $\hat{a}$ are the parameters that optimize the likelihood for a given model $M$, $k(a)$ is the number of parameters and $N$ the number of data points. The $BIC$ becomes smaller with better model fit. In addition, Table \ref{tab:lmm_modcomp} gives the deviance, $-2 \log p(\mathbf{x}| M, \hat{a})$, which is higher for better model fit, and the $\chi^2$ with the corresponding $p$-value, describing the likelihood ratio, which follows a $\chi^2$-distribution.

\begin{table}[!h]
	\caption{Results of parametric statistical testing using model comparison between linear mixed models. Simple effects of factors \textit{isoflurane level} and \textit{direction} or \textit{recording site} were tested against the null model; interaction effects  \textit{isoflurane level} $times$ \textit{direction} and \textit{isoflurane level} $times$ \textit{recording site} were tested against the additive models \textit{isoflurane level} $+$ \textit{direction} and \textit{isoflurane level} $+$ \textit{recording site}, respectively. *** as defined in table \ref{tab:pANOVA}}
	\label{tab:lmm_modcomp}
    \begin{adjustbox}{center}
	\begin{tabular}{lllccccc}
		\hline
		\textbf{animal} & \textbf{measure} & \textbf{effect} & \textbf{BIC} & \textbf{deviance} & $\mathcal{X}^2$  & $\mathcal{X}$ df & $p$ \\ \hline

		Ferret 1
		& $TE_{SPO}$  & \textit{null model}   & 345430.8 &  345401.7 & NA & NA & NA \\
		&             & \textit{isoflurane level}   & 345421.5 &  345373.1 &   28.58 & 2 & $<<$0.000$^{***}$ \\
		&             & \textit{direction}    & 342828.4 &  342789.7 & 2612.04 & 1 & $<<$0.000$^{***}$ \\ 
		&             & \textit{isoflurane l.} + \textit{direction}  & 342819.3 &  342761.3 & NA & NA & NA \\
		&             & \textit{isoflurane l.} $\times$ \textit{direction} & 342139.9 &  342062.4 & 698.84 & 2 & $<<$0.000$^{***}$ \\ \hline

        & $AIS$       & \textit{null model}     & 37581.2 &  37551.8 & NA & NA & NA \\
        &             & \textit{isoflurane level}     & 37555.8 &  37506.8 &   46.00 & 2 & $<<$0.000$^{***}$ \\
        &             & \textit{recording site} & 33296.2 &  33256.9 & 4294.82 & 1 & $<<$0.000$^{***}$ \\ 
        &             & \textit{isoflurane l.} + \textit{recording site}  & 33270.9 &  33212.1 & NA & NA & NA \\
        &             & \textit{isoflurane l.} $\times$ \textit{recording site} & 30754.3 &  30675.8 & 2536.29 & 2 & $<<$0.000$^{***}$ \\ \hline

        & $H$         & \textit{null model}     & 353650.8 &  353621.8 & NA & NA & NA \\
        &             & \textit{isoflurane level}     & 353653.1 &  353604.7 &   17.03 & 2 & \textless0.000$^{***}$ \\
        &             & \textit{recording site} & 353446.8 &  353408.1 &  213.64 & 1 & $<<$0.000$^{***}$ \\ 
        &             & \textit{isoflurane l.} + \textit{recording site}  & 353449.1 &  353391.1 & NA & NA & NA \\
        &             & \textit{isoflurane l.} $\times$ \textit{recording site} & 347204.7 &  347127.2 & 6263.85 & 2 & $<<$0.000$^{***}$ \\ \hline

		Ferret 2
		& $TE_{SPO}$  & \textit{null model}   & 131164.7 &  131135.3 & NA & NA & NA \\
		&             & \textit{isoflurane level}   & 131152.3 &  131103.2 &  32.06 & 2 & $<<$0.000$^{***}$ \\
		&             & \textit{direction}    & 131168.9 &  131129.6 &   5.66 & 1 & 0.020 \\ 
		&             & \textit{isoflurane l.} + \textit{direction}  & 131109.5 &  131097.5 & NA & NA & NA \\
		&             & \textit{isoflurane l.} $\times$ \textit{direction} & 130871.5 &  130793.0 & 304.56 & 2 & $<<$0.000$^{***}$ \\ \hline

        & $AIS$       & \textit{null model}     & 109203.1 &  109173.7 & NA & NA & NA \\
        &             & \textit{isoflurane level}     & 109183.1 &  109134.1 &   39.63 & 2 & $<<$0.000$^{***}$ \\
        &             & \textit{recording site} & 105827.2 &  105788.0 & 3385.68 & 1 & $<<$0.000$^{***}$ \\ 
        &             & \textit{isoflurane l.} + \textit{recording site}  & 105807.3 & 105748.4 & NA & NA & NA \\
        &             & \textit{isoflurane l.} $\times$ \textit{recording site} & 104173.5 &  104095.0 & 1653.46 & 2 & $<<$0.000$^{***}$ \\ \hline

        & $H$         & \textit{null model}     & -14074.8 &  -14104.2 & NA & NA & NA \\
        &             & \textit{isoflurane level}     & -14088.6 &  -14137.7 &    33.43 & 2 & \textless0.000$^{***}$ \\
        &             & \textit{recording site} & -18423.0 &  -18462.3 &  4358.04 & 1 & $<<$0.000$^{***}$ \\ 
        &             & \textit{isoflurane l.} + \textit{recording site}  & -18436.7 &  -18495.6 & NA & NA & NA \\
        &             & \textit{isoflurane l.} $\times$ \textit{recording site} & -19867.2 &  -19945.7 & 1450.09 & 2 & $<<$0.000$^{***}$ \\ \hline
	\end{tabular}
    \end{adjustbox}
	\vspace*{-4pt}
\end{table}

\paragraph*{Bayesian estimation on discretized data} In addition to the neighbor-distance based estimators for $TE_{SPO}$, $AIS$, and $H$ used above, we also applied Bayesian estimators  recently proposed by Nemenman, Shafe, and Bialek (NSB) \cite{nsb,nsb2} to our data. The Bayesian approach promises to yield unbiased estimators when priors are chosen appropriately. However, one has to keep in mind that these estimators currently require the discretization of continuous data, and therefore may loose important information.

\begin{figure}[!h]
   \includegraphics{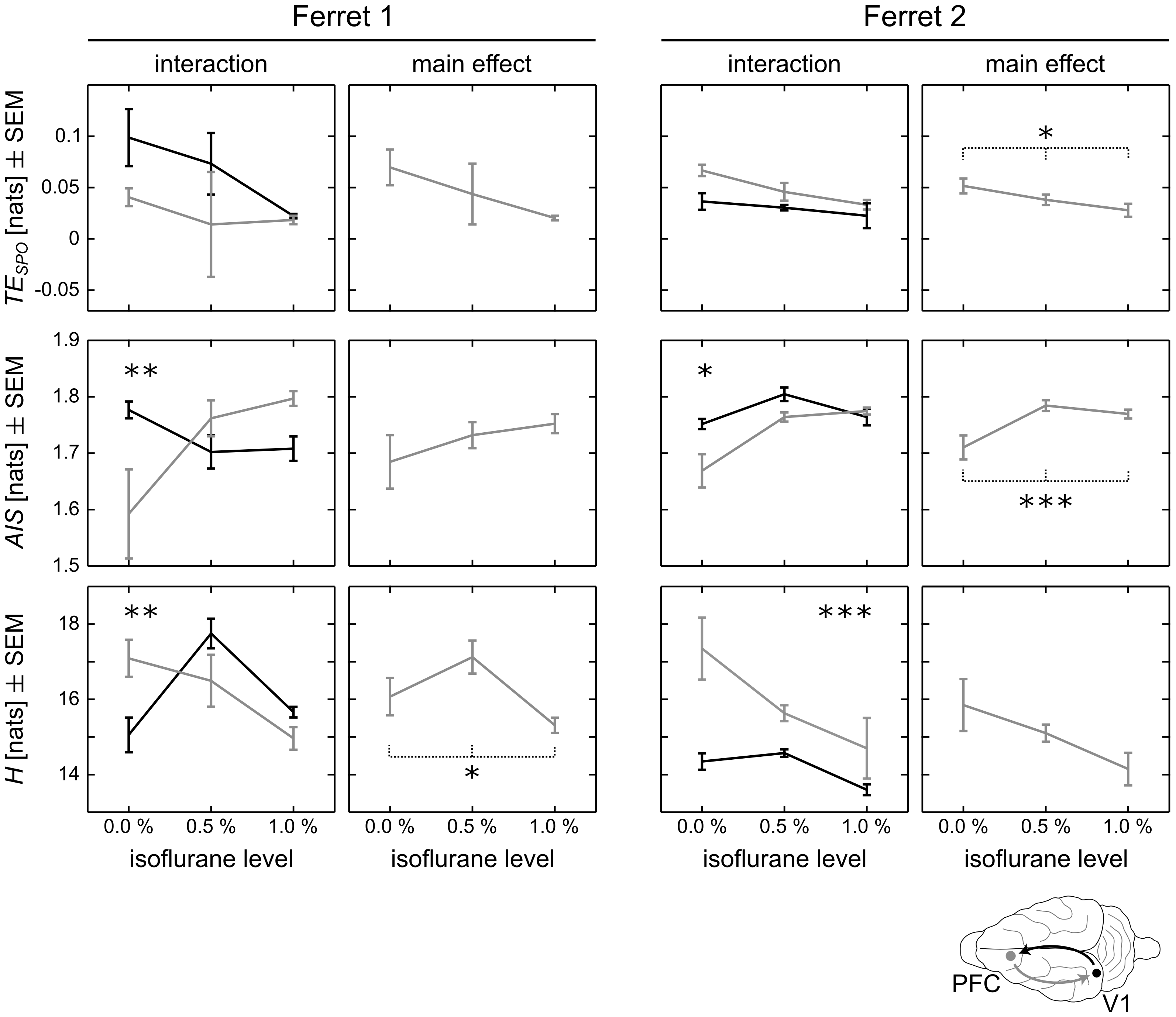}
\caption{{\bf pANOVA results for Bayesian estimates of transfer entropy ($TE_{SPO}$), active information storage ($AIS$), and entropy ($H$).}
Left columns show interactions \textit{isoflurane level x direction} and \textit{isoflurane level x recording site} for both animals; right columns show main effects \textit{isoflurane level}. Grey lines in interaction plots indicate $TE_{SPO}$ from prefrontal cortex (PFC) to primary visual areas (V1), or $H$ and $AIS$ in PFC; black lines indicate $TE_{SPO}$ from V1 to PFC, or $H$ and $AIS$ in V1. Error bars indicate the standard error of the mean (SEM); stars indicate significant interactions or main effects ($^* p<0.05$; $^{**} p<0.01$; $^{***} p<0.001$).}
\label{fig:pANOVA_main_effects_bayes}
\end{figure}

 When applying the NSB estimator to the discretized LFP time series with $N_{bins}$ discretization steps, we observed that for $N_{bins} \geq 8$ the results were qualitatively consistent for different choices of numbers of bins. We present results for $N_{bins}=12$ (Fig. \ref{fig:pANOVA_main_effects_bayes}), which provides a reasonable resolution of the signal while still allowing for a reliable estimation of entropies within the scope of available data.

\begin{figure}[!h]
   \includegraphics{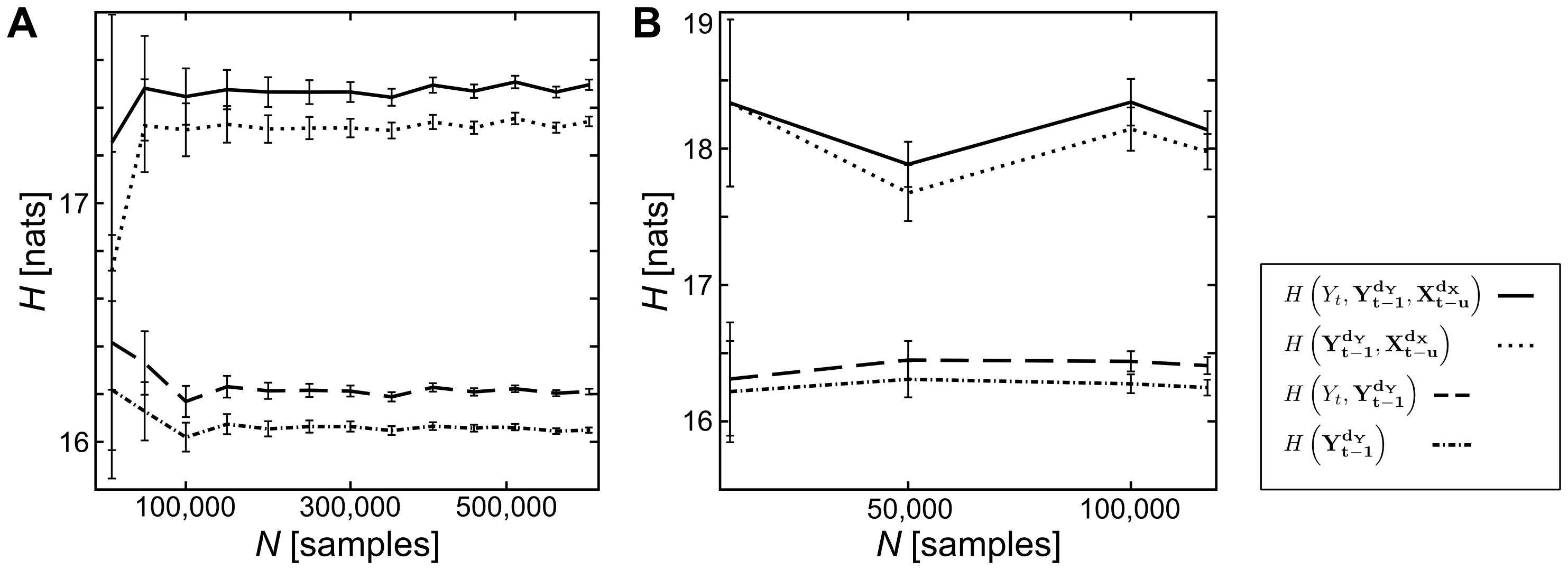}
\caption{{\bf Examples for estimates of entropy ($H$) terms for transfer entropy calculation (Eq. \ref{eq:te_net}) by number of data points $N$, using the Nemenman-Shafee-Bialek-estimator (NSB).}
(A) entropies for ferret 1, V1, \textit{iso 0.5~\%}: estimates are stable for $N\geq 100,000$; (B) entropies for ferret 2, PFC, \textit{awake}: an insufficient number of data points does not allow for verification of the estimate's robustness (recording was excluded from further analysis). Variables labeled $X$ reflect data from the source variable, $Y$ from the target variable. $t$ is an integer time index, and bold typeface indicates the state of a system (see Methods).}
\label{fig:nsb_est_ex}
\end{figure}

We confirmed the reliability of the estimator by systematically reducing the sample size and found no substantial impact on our estimates (Fig. \ref{fig:nsb_est_ex}).

The estimation of $H$, $AIS$, and $TE_{SPO}$ by Bayesian techniques for the binned signal representations provided results that were qualitatively consistent with results from the neighbor-distance based estimation techniques (compare Figs. \ref{fig:pANOVA_main_effects_all} and \ref{fig:pANOVA_main_effects_bayes}, and Tables \ref{tab:pANOVA} and \ref{tab:permANOVA_bayes}). While the Bayesian estimates showed larger variances across different recordings and sample sizes, on average, we saw a decrease of $TE_{SPO}$ and $H$ for higher concentrations of isoflurane (main effect \textit{isoflurane level}), while $AIS$ increased for higher concentrations. For ferret 1 we also found an interaction effect for $TE_{SPO}$, with a stronger reduction in information transfer in top-down direction.

\begin{table}[!h]
	\caption{Results of permutation analysis of variance for information theoretic measures obtained through Bayesian estimation ($p$-values).}
	\label{tab:permANOVA_bayes}
	\begin{tabular}{llcc}
		\hline
		\textbf{measure} & \textbf{effect} & \textbf{Ferret 1} & \textbf{Ferret 2} \\ \hline

		$TE_{SPO}$  & \textit{isoflurane level} & 0.0549  & 0.0013$^{*}$ \\
		& \textit{direction} & 0.0272$^{*}$ &  0.0002$^{***}$ \\
		& \textit{interaction} & 0.6627 & 0.0616 \\ \hline

		$AIS$  & \textit{isoflurane level} & 0.2148  & \textless0.0001$^{***}$ \\
		& \textit{direction} & 0.0788 &  0.0032$^{**}$ \\
		& \textit{interaction} & 0.0026$^{**}$ & 0.0153$^{*}$ \\ \hline

		$H$   & \textit{isoflurane level} &  0.0184$^{*}$ & \textless0.0001$^{***}$ \\
		& \textit{recording site} & 0.2738 &  \textless0.0001$^{***}$ \\
		& \textit{interaction} & 0.0017$^{ **}$  & 0.0053 \\ \hline
		\multicolumn{4}{l}{$^* p<0.05$; $^{**} p<0.01$; $^{***} p<0.001$} \\
	\end{tabular}
	\vspace*{-4pt}
\end{table}

Note that we also applied an alternative Bayesian estimation scheme based on Pitman-Yor-process priors \cite{archer2014}. However, for this estimation procedure, we observed that the data were insufficient to allow for a robust estimation of the tailing behavior of the distribution as indicated by large variances and unreasonably high estimates across different sample sizes.

\paragraph*{Simulated effects of changed source entropy on transfer entropy} To test the effect of reduced source entropy on $TE_{SPO}$ between a source and target process, we simulated two test cases with high and low source entropy respectively. In both test cases signals were based on the original recordings, and the coupling between source and target was held constant (see Methods).

\begin{figure}[!h]
   \includegraphics{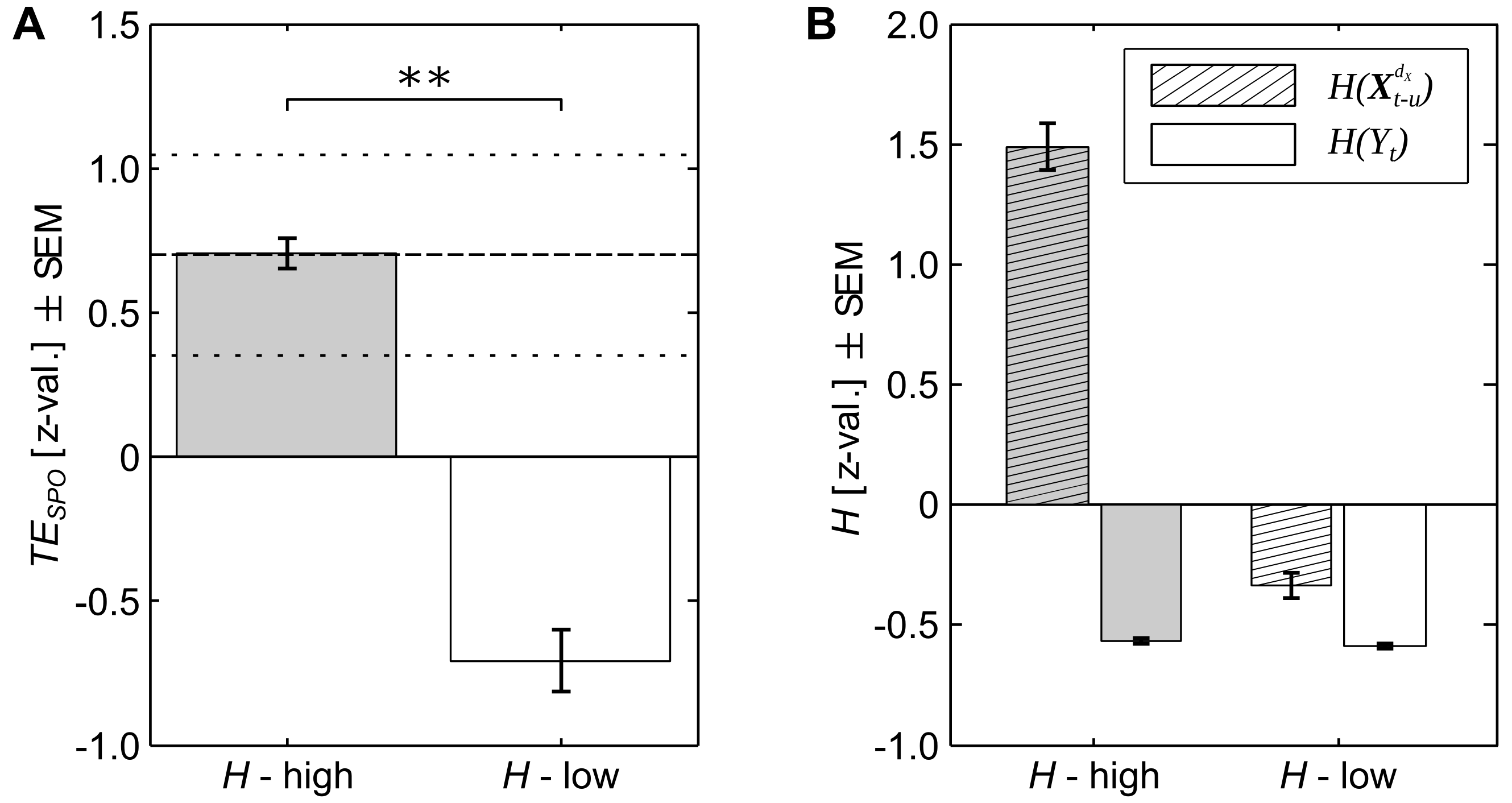}
    \caption{{\bf Simulated effects of changed source entropy on transfer entropy ($TE_{SPO}$).}
    (A) $TE_{SPO}$ for two simulated cases of high (\textit{$H$-high}) and low (\textit{$H$-low}) source entropy ($^{**} p<0.01$, error bars indicate the standard error of the mean, SEM, over data epochs). Dashed lines indicate the average $TE_{SPO}$ estimated from the original data under 0.0~\% isoflurane $\pm$ SEM;
    (B) source entropy $H(\mathbf{X}_{t-u}^{d_X})$ (dashed bars) and target entropy $H(Y_{t})$ (empty bars) for the two simulated test cases of high (gray bars) and low entropy (white bars), error bars indicate the SEM over data epochs. Source entropy was higher in the high-entropy simulation, while target entropy was approximately the same for both cases. Results are given as z-values across estimates for all epochs from both simulations.
    }
\label{fig:simulation_entropy}
\end{figure}

We found significantly higher $TE_{SPO}$ in the case with high source entropy than in the case with low source entropy (Fig. \ref{fig:simulation_entropy}A). The simulation thus demonstrated that a lower source entropy does indeed lead to a reduction in $TE_{SPO}$ despite an unchanged coupling. Information transfer in the high-entropy case was similar to the average information transfer found in recordings under an isoflurane concentration of 0.0~\%, indicating that the simulation scheme mirrored information transfer found in the experimental recordings.

\subsection*{Optimized embedding parameters for $TE_{SPO}$ and $AIS$ estimation}
As noted in the introduction, the estimation of information theoretic measures from finite data is challenging. For the measures that describe distributed computation in complex systems, such as transfer entropy, estimation is further complicated because the available data are typically only scalar observations of a process with multidimensional dynamics. This necessitates the approximate reconstructions of the process' states via a form of embedding \cite{ragwitz2002}, where a certain number of past values of the scalar observations spaced by an embedding delay are taken into account (e.g. for a pendulum swinging through its zero position one additional past position values will clarify whether it's going left or right). An important part of proper transfer entropy estimation is thus optimization of this number of past values (embedding dimension), and of the embedding delay. These two embedding parameters then approximately define past states, whose correct identification is crucial for the estimation of transfer entropy \cite{vicente2011TE}, but also for the estimation of active information storage.  Without it information storage may be underestimated and, erroneous values of the information transfer will be obtained; even a detection of information transfer in the wrong direction is likely (see \cite{vicente2011TE,lindner2011,wibral2013timing} and Methods section).
Existing studies using transfer entropy often omitted the optimization of embedding parameters, and instead used ad-hoc choices, which may have had a detrimental effect on the estimation of transfer entropy---hence the need for a confirmation of previous results in this study.

In the present study, we therefore used a formal criterion proposed by Ragwitz \cite{ragwitz2002}, to find an optimal embedding defined by the embedding dimension $d$ (the number of values collected) and delay $\tau$ (the temporal spacing between them), to find embeddings for the $TE_{SPO}$ and $AIS$ past states. We used the implementation of this criterion provided by the TRENTOOL toolbox. Since the bias of the estimators used depends on $d$, we used the maximum $d$ over all conditions and directions of interaction as the common embedding dimension for estimation from each epoch to make the estimated values statistically comparable. The resulting dimension used was 15 samples, which is considerably higher than the value commonly used in the literature \cite{lee2013,ku2011STE}, when choosing the embedding dimension ad-hoc or using other criteria as for example the sample size \cite{untergehrer2014}. The embedding delay $\tau$ was optimized individually for single epochs in each condition and animal as it has no influence on the estimator bias.

\subsection*{Relevance of individual information transfer delay estimation}
Several previous studies on information transfer under anesthesia reported the sign of the so called net transfer entropy ($TE_{net}$) as a measure of the predominant direction of information transfer between two sources. $TE_{net}$ is essentially just the normalized difference between the transfer entropies measured in the two direction connecting a pair of recording sites (see Methods). When calculating $TE_{net}$, it is particularly important to individually account for physical delays $\delta$ in the two potential directions of information transfer, because otherwise the sign of $TE_{net}$ may become meaningless (see examples in \cite{wibral2013timing}). As this delay is unknown a priori it has to be found prior to the actual estimation of information transfer. We have recently shown that this can be done by using a variable delay parameter $u$ in a delay sensitive transfer entropy estimator $TE_{SPO}$ (see Methods); here, the $u$ that maximizes the transfer entropy over a range of assumed values for $u$ reflects the physical delay \cite{wibral2013timing}.

\begin{figure}[!h]
   \includegraphics{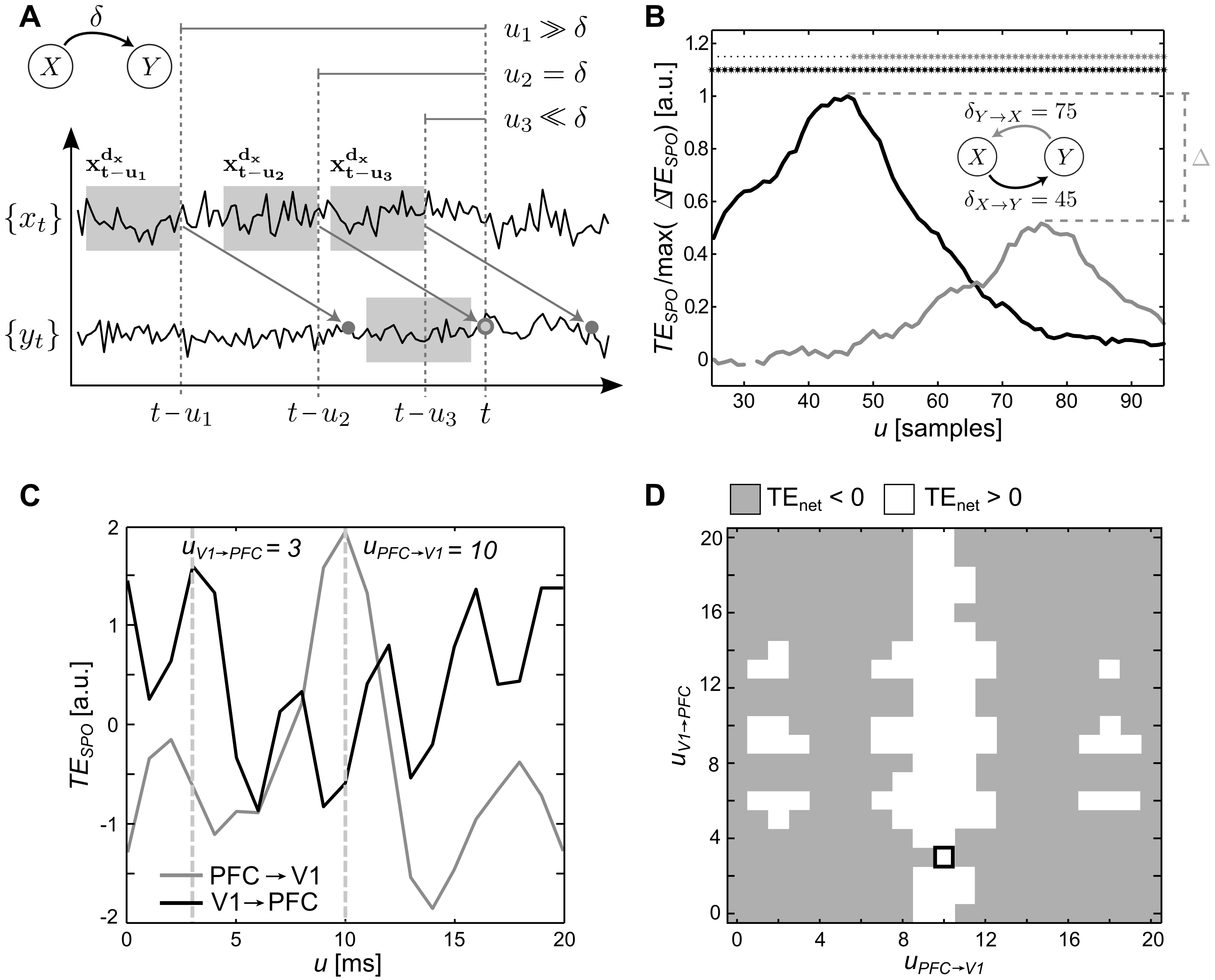}
\caption{{\bf Estimation of information transfer delays.}
	(A, modified from \protect\cite{wibral2013timing}) estimation of transfer entropy ($TE_{SPO}$) depends on the choice of the delay parameter $u$, if $u$ is much smaller or bigger than the true delay $\delta$, information arrives too late or too early in the target time series and information transfer is not correctly measured;
        (B, modified from \protect\cite{wibral2013timing}) $TE_{SPO}$ values estimated from two simulated, bidirectionally coupled Lorenz systems (see \protect\cite{wibral2013timing} for details) as a function of $u$ for both directions of analysis, $TE_{SPO}(X \rightarrow Y,t,u)$ (black line) and $TE_{SPO}(Y \rightarrow X,t,u)$ (gray line): $TE_{SPO}$ values vary with the choice of $u$ and so does the absolute difference between values; using individually optimized transfer delays for both directions of analysis yields the meaningful difference $\Delta$ (dashed lines), where $TE_{SPO}(X\rightarrow Y,t,u_{opt}) > TE_{SPO}(Y\rightarrow X,t,u_{opt})$;
        (C) example transfer entropy analysis for one recorded epoch: $TE_{SPO}$ values vary greatly as a function of $u$, optimal choices of $u$ are marked by dashed lines;
        (D) sign (gray: negative, white: positive) of $TE_{net}$ for different values of $u_{PFC \rightarrow V1}$ (x-axis) and $u_{V1 \rightarrow PFC}$ (y-axis), calculated from $TE_{SPO}$ values shown in panel C: the sign varies with individual choices of $u$; the black frame marks the combination of individually optimal choices for both parameters that yields the correct result.}
\label{fig:delay_reconstruction}
\end{figure}

The necessity to individually optimize $u$ for each interaction is not a mere theoretical concern but was clearly visible in the present study: Fig. \ref{fig:delay_reconstruction}C shows representative results from ferret 1 under 0.0~\% isoflurane, where apparent $TE_{SPO}$ values strongly varied as a function of the delay $u$. As a consequence, $TE_{net}$ values also varied if a \textit{common} delay $u$ was chosen for both directions. In other words, the sign of the $TE_{net}$ varied as a function of individual choices for $u_{PFC \rightarrow V1}$ and $u_{V1 \rightarrow PFC}$ for each direction of information transfer and hence became uninterpretable (Fig. \ref{fig:delay_reconstruction}D).

\begin{figure}[!h]
   \includegraphics{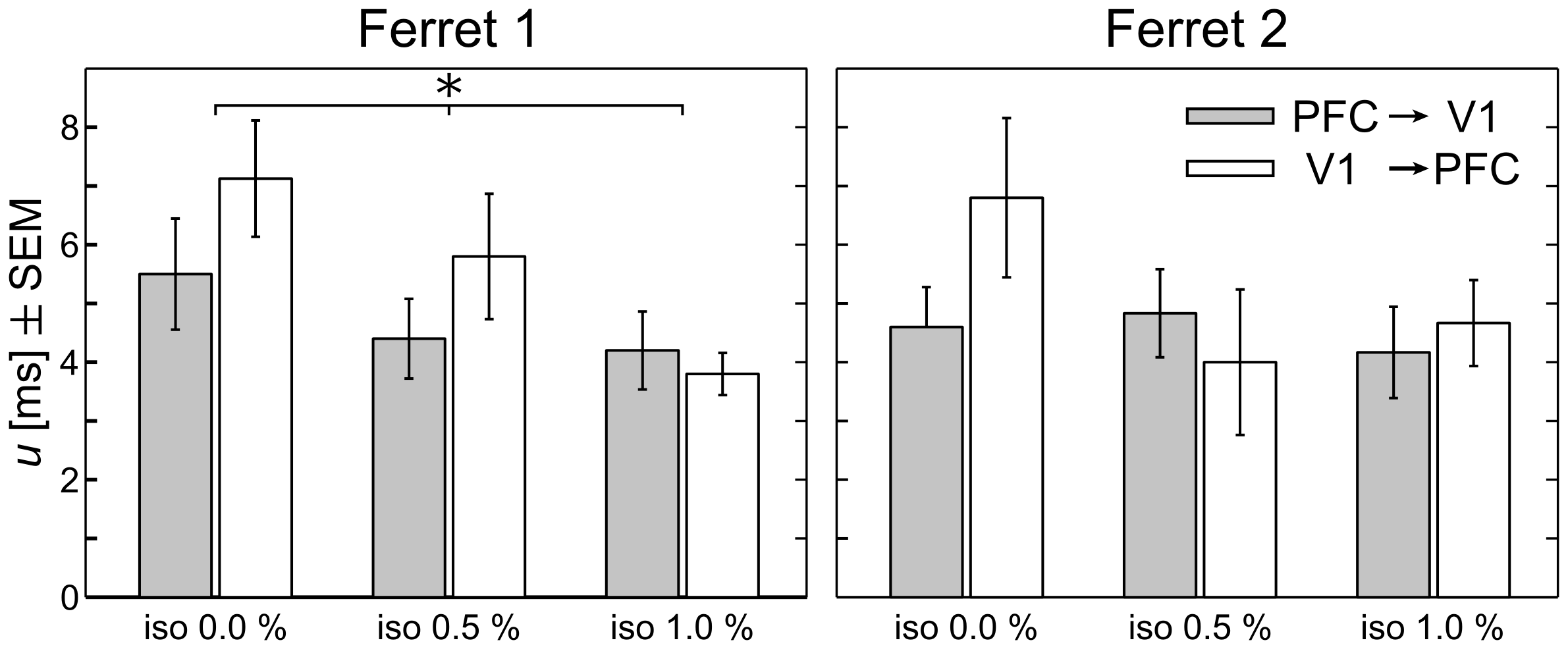}
\caption{{\bf Optimized information transfer delays $u$ for both directions of interaction and three levels of isoflurane, by animal.}
Bars denote averages over recordings per condition; error bars indicate the standard error of the mean (SEM). There was a significant main effect of \textit{isoflurane level} for ferret 1 ($p < 0.05$).}
\label{fig:opt_u}
\end{figure}

As a consequence of the above, we here individually optimized $u$ for each direction of information transfer in each condition and animal to estimate the true delay of information transfer following the mathematical proof in \cite{wibral2013timing}. We individually optimized $u$ to obtain estimates of transfer entropy that were not biased by a non-optimal choice for $u$. We used the implementation in TRENTOOL \cite{lindner2011} and scanned values for $u$ ranging from 0 to 20~ms. Averages for optimized delays ranged from 4-7~ms across animals and isoflurane levels (Fig. \ref{fig:opt_u}).

\begin{figure}[!h]
    \includegraphics{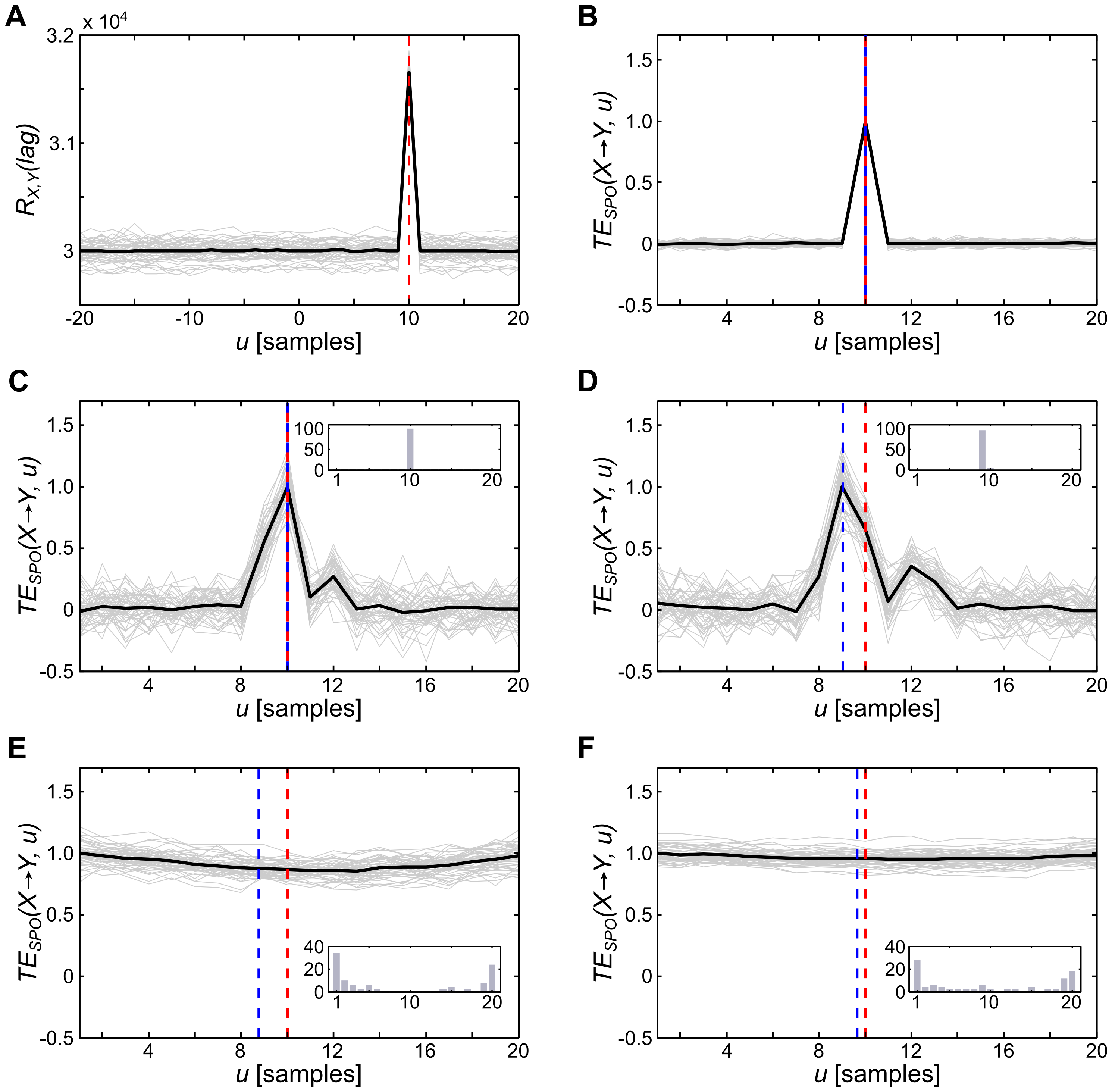}
    \caption{{\bf Simulated effects of filtering on the reconstruction of information transfer delays ($u$).}
    (A) Cross-correlation $R_{X,Y}$ between two simulated, coupled time series ($N=100000$, drawn from a uniform random-distribution over the open interval $(0,1)$, true coupling delay 10~samples, indicated by the red dashed line), the simulation was run 50 times, black lines indicate the mean over simulation runs, gray lines indicate individual runs;
    (B) $TE_{SPO}$ as a function of information transfer delay $u$ before filtering the data;
    (C--D) $TE_{SPO}$ as a function of $u$ after filtering the data with a bandpass filter (fourth order, causal Butterworth filter, implemented in the MATLAB toolbox FieldTrip \protect\cite{oostenveld2011FT}); red dashed lines indicate the simulated delay $\delta$, blue lines indicate the average reconstructed delay $u$ over simulation runs, histograms (inserts) show the distribution of reconstructed values for $u$ over simulation runs in percent;
    (C) broadband filtering (0.1--300~Hz) introduced additional peaks in $TE_{SPO}$ values, however, the maximum peak indicating the optimal $u$ was still at the simulated delay for all simulated runs;
    (D) filtering within a narrower band (0.1--200~Hz) led to an imprecise reconstruction of the correct $ \delta$ in each run with an error of 1 sample, i.e. $\delta$ tended to be underestimated;
    (E--F) narrow-band filtering in the beta range (12--30~Hz) and theta range (4--8~Hz) led to a wide distribution of $u$ with a large absolute error of up to 10~samples.
    }
\label{fig:simulation_filtering_delay}
\end{figure}

Note that in Fig. \ref{fig:delay_reconstruction}C, $TE_{SPO}$ as a function of $u$ shows multiple peaks, especially for the direction $V1 \rightarrow PFC$. Since these peaks resembled an oscillatory pattern close to the low-pass filter frequency used for preprocessing the data, we investigated the influence of filtering on delay reconstruction by simulating two coupled time series for which we reconstructed the delay with and without prior band-pass filtering (Fig. \ref{fig:simulation_filtering_delay}). For filtered, simulated data $TE_{SPO}$ as a function of $u$  indeed showed an oscillatory pattern for certain filter settings. However, broadband filtering---as used here for the original data---did not lead to the reconstruction of an incorrect value for the delay $\delta$ (Fig. \ref{fig:simulation_filtering_delay}C). Yet, when filtering using narrower bandwidths, the reconstruction of the correct information transfer delay failed (Fig. \ref{fig:simulation_filtering_delay}D--F). This finding supports the general notion that filtering, especially narrow-band filtering, may only be applied with caution before estimating connectivity measures \cite{barnett2011,florin2010}.

Yet, since broad-band filtering did not lead to the observed oscillatory patterns in the simulated data, alternative explanations for multiple peaks in $TE_{SPO}$ are more likely: one alternative cause of multiple peaks in the $TE_{SPO}$ are multiple information transfer channels with various delays between source and target (see \cite{wibral2013timing}, especially Fig. 6). Each information transfer channel with its individual delay will be detected when estimating $TE_{SPO}$ as a function of $u$ (see simulations in \cite{wibral2013timing}). A further possible cause for multiple peaks in the $TE_{SPO}$ is the existence of a feedback loop between the source of $TE_{SPO}$ and a third source of neural activity \cite{wibral2013timing}. In such a feedback loop, information in the source is fed back to the source such that it occurs again at a later point in time, leading to the repeated transfer of identical bits of information. Both causes are potential explanations for the occurrence of multiple peaks in the $TE_{SPO}$ in the present study---yet, deciding between these potential causes requires additional research, specifically interventional approaches to the causal structure underlying the observed information transfer.

In sum, our results clearly indicate the necessity to individually optimize information transfer delays and that often employed ad-hoc choices for $u$ may result in spurious results in information transfer.The additional simulations of filtering effects showed that narrow-band filtering may have detrimental effects on this optimization procedure and should thus be avoided.

\subsection*{Relation of information-theoretic measures with other time-series properties}
\begin{figure}[!h]
   \begin{adjustbox}{center}
   \includegraphics{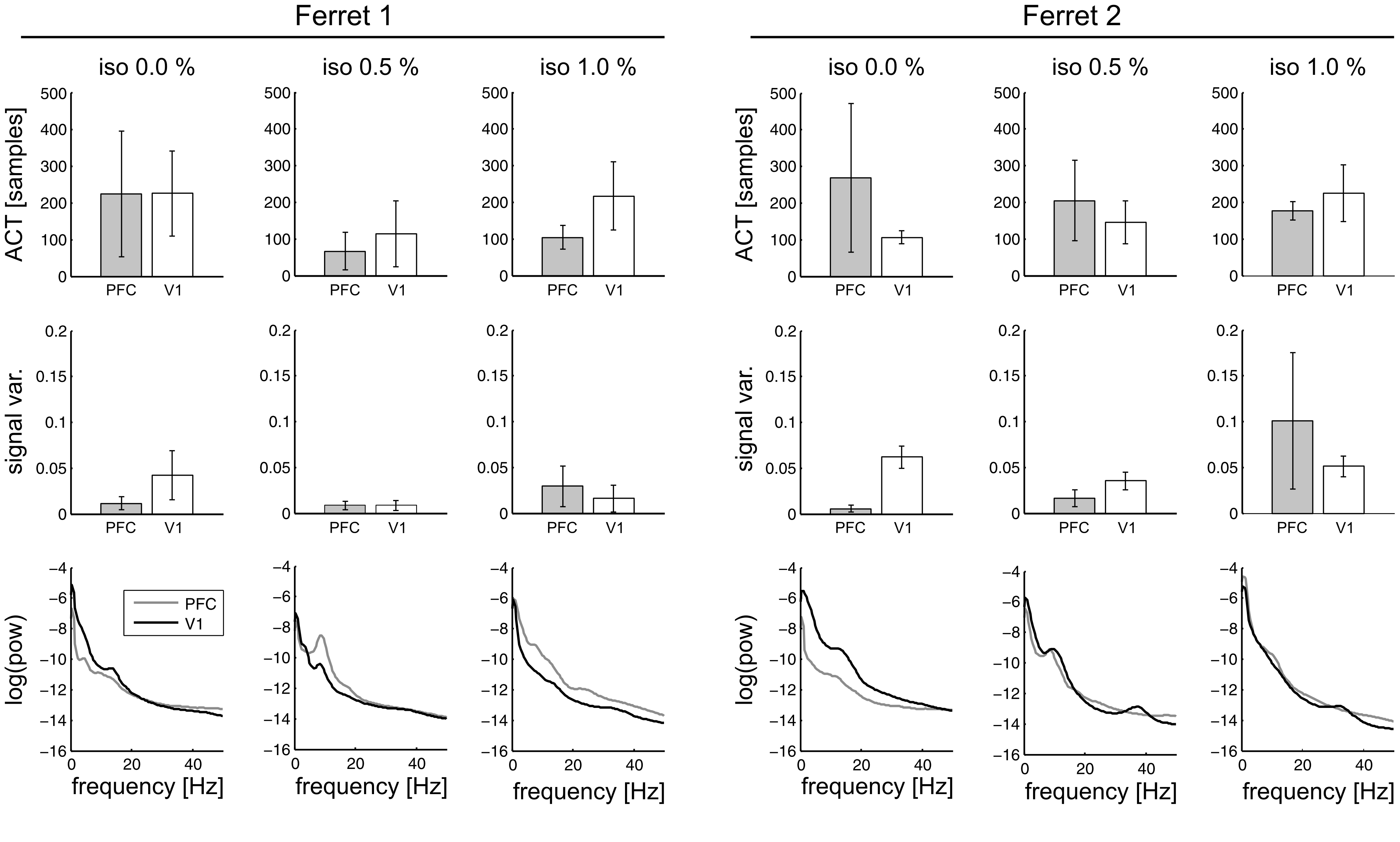}
   \end{adjustbox}
	\caption{{\bf Autocorrelation decay time (ACT), signal variance, and power by isoflurane level and recording site for both animals.}
	Averages over recordings per isoflurane level and recording site; error bars indicate one standard deviation.}
\label{fig:corr_raw_values}
\end{figure}

The information-theoretic measures used in this study are relatively novel and have been applied in neuroscience research rarely. Thus, it is conceivable that these measures quantify signal properties that are more easily captured by traditional time-series measures, namely, the autocorrelation decay time (ACT), the signal variance, and the power in individual frequency bands. To investigate the overlap between information-theoretic and traditional measures, we calculated correlations between $AIS$ and ACT, between $H$ and signal variance, between $TE_{SPO}$ and ACT, between $TE_{SPO}$ and signal variance, and between $AIS$, $H$, and $TE_{SPO}$ and the power in various frequency bands, respectively. Fig. \ref{fig:corr_raw_values} shows average ACT, signal variance, and power over recordings, for individual isoflurane levels and recording sites, and for both animals.

Detailed tables showing correlation coefficients and explained variances are provided as supporting information S2. In summary, for $AIS$ and ACT we found significant correlations in both animals, however the median of the variance explained, $\widetilde{R}^2$, over individual recordings was below 0.1  for each significant correlation in an animal, recording site and isoflurane level. Hence, even though there was some shared variance between ACT and $AIS$, there remained a substantial amount of unexplained variance in $AIS$, indicating that $AIS$ quantified properties other than the signal's ACT---in line with results from our earlier studies \cite{gomez2014,brodski2016}. This is expected because the autocorrelation (decay) time ACT measures how long information persists in a time series in a linear encoding, while $AIS$, in contrast, measures how much information is stored (at any moment), and also reflects nonlinear transformations of this information.
The fact that there is still some shared variance between the two measures here may be explained by the construction of the $AIS$'s past state embedding, where the time steps between the samples in the embedding vector are defined as a fractions of the ACT.

Between $H$ and signal variance, we found no significant correlation; moreover, correlations for individual recordings were predominantly negative.

The $TE_{SPO}$ was significantly correlated with source ACT for an isoflurane level of 1.0~\%. Also for the correlation of $TE_{SPO}$ and the source's variance, we found significant correlations for the bottom-up direction under 0.5~\% and 1.0~\% isoflurane in animal 1, and for both directions under 1.0~\% isoflurane in animal 2. For all significant correlations we found $\widetilde{R}^2 \leq 0.015$, again indicating a substantial amount of variance in $TE_{SPO}$ that was not explained by ACT or signal variance.

When correlating band power with information-theoretic measures, we found significant correlations between $AIS$ and all bands in both animals animals (see table \ref{tab:ais_band_pow} as part of supporting information S2, $\widetilde{R}^2 < 0.16$); for $H$ and band power, we found correlations predominantly in higher frequency bands (beta and gamma, see table \ref{tab:h_band_pow} as part of  supporting information S2, $\widetilde{R}^2 < 0.1$); for $TE_{SPO}$ and band power, we found significant correlations predominantly in the gamma band and one significant correlation in the delta band for animal 2 and in the beta band for animal 1, respectively (see table \ref{tab:te_band_pow} as part of supporting information S2, $\widetilde{R}^2 < 0.024$). In sum, we found relationships between the power in individual frequency bands and all three information-theoretic measures. For all measures, the variance explained was below 0.2, indicating again that band power did not fully capture the properties measured by $TE_{SPO}$, $AIS$, and $H$.

\section*{Discussion}

We analyzed long-range information transfer between areas V1 and PFC in two ferrets under different levels of isoflurane. We found that transfer entropy was indeed reduced under isoflurane and that this reduction was more pronounced in top-down directions. These results validate earlier findings made using different estimation procedures \cite{imas2005,jordan2013,ku2011STE,lee2013,untergehrer2014}. As far as information transfer alone was concerned our results are compatible with an interpretation of reduced long-range information transfer due to reduced inter-areal coupling. Yet, this interpretation provides no direct explanation for the findings of reduced locally available information as explained below. In contrast, the alternative hypothesis that the reduced long range information transfer is a secondary effect of changes in local information processing provides a concise explanation for our findings both with respect to locally available information, and information transfer.

\subsection*{Reduction in transfer entropy may be caused by changes in local information processing}
To test our alternative hypothesis we evaluated two of its predictions about changes in locally available information, as measured by signal entropy, under administration of different isoflurane concentrations. First, entropy should be reduced; second, the strongest decrease in information transfer should originate from the source node with the strongest decrease in entropy, rather than end in this node. Indeed, we found that signal entropy decreased; the most pronounced decrease in signal entropy was found in PFC. In accordance with our prediction, we found that PFC---the node with the larger decrease in entropy---was also at the source, not the target, of the most pronounced decrease in transfer entropy. This is strong evidence against the theoretical possibility that in a recording site, entropy decreased due to a reduced influx of information---because in this latter scenario the strongest reduction in entropy should have been found at the target (end point) of the most pronounced decrease in information transfer. Hence, our hypothesis that long-range cortico-cortical information transfer is reduced because of changes in local processing must be taken as a serious alternative to the currently prevailing theories of anesthetic action based on disruptions of long range interactions. We suggest that a renewed focus on local information processing in anesthesia research will be pivotal to advance our understanding of how consciousness is lost.

Our predictions that reductions in entropy should potentially be reflected in reduced information transfer derives from the simple principle that information that is not available at the source cannot be transferred. We may thus in principle reduce $TE_{SPO}$ to arbitrarily low values by reducing the entropy of one of the involved processes, without changing the physical coupling between the two systems, just by changing their internal information processing. This trivial but important fact has been neglected in previous studies when interpreting changes in information transfer as changes in coupling strength. Even when this bound is not attained, e.g. because only a certain fraction of the local information is transferred even under 0.0~\% isoflurane, it seems highly plausible that reductions in the locally available information affect the amount of information transfered. This line of reasoning generalizes to applications beyond anesthesia research, i.e., every application that observes changes in information transfer between two or more experimental conditions should consider changes in local information processing as a potential alternative cause.

A possible indication of how exactly local information processing is changed is given by the observation of increased active information storage in PFC and V1 (also see the next paragraph). This means that more 'old' information is kept stored in a source's activity under anesthesia, rather than being dynamically generated. Such  stored source information will not contribute to a measurable transfer entropy under most circumstances  because it is already known at the target (see \cite{wibral2015a}, section 5.2.3, for an illustrative example).

\subsection*{Relation between changed locally available information and information storage}
In general, $AIS$ increases if a signal becomes more predictable when knowing it's past, but is unpredictable otherwise.  This also means that the absolute $AIS$ is upper-bounded by the system's entropy $H$ (see Methods, Eq. \ref{eq:ais_entropy}). Thus, a decrease in $H$ can in principle lead to a decrease in $AIS$, i.e., fewer possible system states may lead to a decrease in absolute $AIS$. However, in the present study, we observed an \textit{increase} in $AIS$ while $H$ decreased---this indicates an increase in predictability that more than compensates for the decrease in locally available information. In other words, the system visited fewer states in total but the next state visited became more predictable from the system's past. Thus, a reduction in entropy and increase in predictability points at highly regular neural activity for higher isoflurane concentrations. Such a behavior in activity is in line with existing electrophysiological findings: under anesthesia signals have been reported to become more uniform, exhibiting repetitive patterns, interrupted by bursting activity (see \cite{alkire2008} for a review). For example, Purdon and colleagues observed a reduction of the median frequency and an overall shift towards high-power, low-frequency activity during LOC \cite{purdon2013}. In particular, slow-wave oscillatory power was more pronounced during anesthesia-induced LOC. LOC was also accompanied by a significant increase in power and a narrowing of oscillatory bands in the alpha frequency range in their study.

\subsection*{Limitations of the applied estimators and measures}

Unfortunately a quantitative comparison between different information theoretic estimates that would directly relate $H$, $AIS$ and $TE_{SPO}$ is not possible using the continuous estimators applied in this study. Their estimates are not comparable because the bias properties of each estimator differ. For each estimator, the bias depends on the number of points used for estimation as well as on the dimensionality of involved variables---however, the exact functional relationship between these two quantities and the bias is unknown and may differ between estimators. (In our application, the dimensionality is determined mainly by the dimension of the past state vectors, and by how many different state variables enter the computation of a measure).

This lack of comparability makes it impossible to normalize estimates; for example, transfer entropy is often normalized by the conditional entropy of the present target state to compare the fraction of transferred information to the fraction of stored information. We forgo this possibility of comparison for a greater sensitivity and specificity in the detection of changes in the individual information theoretic measures here. New estimators, e.g. Bayesian estimators like the ones tested here, promise more comparable estimates by tightly controlling the biases. Yet, these estimators were not as reliable as expected in our study, displaying a relatively high variance.

A further important point to consider when estimating transfer entropy between recordings from neural sites, is the possibility of third unobserved sources influencing the information transfer. In the present study, third sources (e.g., in the thalamus) may influence the information transfer between PFC and V1---for example, if a third source drove the dynamics in both areas, the areas would become correlated, leading to non-zero estimates of $TE_{SPO}$. This $TE_{SPO}$ is then attributable to the correlation between source and target, but does not measure an actual information transfer between sources. Hence, information transfer estimated by transfer entropy should in general not be directly equated with a causal connection or causal mechanism existing between the source and target process (see also \cite{lizier2010} for a detailed discussion of the difference between transfer entropy and measures of causal interactions).

\subsection*{Potential physiological causes for altered information transfer under anesthesia}

We here tested the possibility that changes in local information processing lead to the frequently observed reduction in information transfer between cortical areas under isoflurane administration, instead of altered long-range coupling. Results on entropies and active information storage suggest that this is a definite possibility from a mathematical point of view.

This is supported by the neurophysiology related to the mode of action of isoflurane, because a dominant influence of altered long-range coupling on $TE_{SPO}$ would mandate that synaptic terminals of the axons mediating long range connectivity should be targets of isoflurane. Such long range connectivity is thought to be dominated by glutamatergic AMPA receptors for inter-areal bottom-up connections, and  glutamatergic NMDA receptors  for inter-areal top-down connections, building upon findings in \cite{felleman1991} and \cite{salin1995} (but see \cite{tomioka2005} for some evidence of GABAergic long range connectivity). Yet, evidence for isoflurane effects on AMPA and NMDA receptors is sparse to date (table 2 in \cite{krasowski1999}). In contrast, the receptors most strongly influenced by isoflurane seem to be $GABA_A$ and nicotinic acetylcholine (nAChR) receptors. More specifically, isoflurane potentiates agonist interactions at the former, while inhibiting the latter.

Thus, if one adopts the current state of knowledge on the synapses involved in long-range inter-areal connectivity, evidence speaks against a dominant effect of modulation of effective long-range connections by isoflurane. This, in turn, points at local information processing as a more likely reason for changed transfer entropy under isoflurane anesthesia. This interpretation is perfectly in line with our finding that decreases in source entropy seem to determine the transfer entropy decreases, instead of  decreases in transfer entropy determining the target entropies.

Nevertheless, targeted local interventions by electrical or optogenetic activation of projection neurons, combined with the set of information theoretic analyses used here, will most likely be necessary to reach final conclusions on the causal role of local entropy changes in reductions of transfered information.

\subsection*{How may altered long-range information transfer lead to loss of consciousness?}

Investigating long-range information transfer under anesthesia is motivated by the question how changed information transfer may cause LOC. To that effect, our findings---a dominant decrease in top-down information transfer under anesthesia, and a decrease in locally available information possibly driving it---may be interpreted in the framework of predictive coding theory \cite{clark2013, hohwy2013predictivemind, hawkins2007}. Predictive coding proposes that the brain learns about the world by constructing and maintaining an internal model of the world, such that it is able to \textit{predict} future sensory input at lower levels of the cortical hierarchy. Whether predictions match actual future input is then used to further refine the internal model. It is thus assumed that top-down information transfer serves the propagation of predictions \cite{bastos2012}. Theories of conscious perception within this predictive coding framework propose that conscious perception is ``determined'' by the internal prediction (or 'hypothesis') that matches the actual input best \cite[p. 201]{hohwy2013predictivemind}. It may be conversely assumed that the absence of predictions leads to an absence of conscious perception.

In the framework of predictive coding theory two possible mechanisms for LOC can be inferred from our data: (1) the disruption of information transfer, predominantly in top-down direction, may indicate a failure to propagate predictions to hierarchically lower areas; (2) the decrease in locally available information and in entropy rates in PFC may indicate a failure to integrate information in an area central to the generation of a coherent model of the world and the generation of the corresponding predictions. These hypotheses are in line with findings reviewed in \cite{hudetz2006} and \cite{mashour2014}, which discuss activity in frontal areas and top-down modulatory activity as important to conscious perception.

Future research should investigate top-down information transfer more closely; for example, recent work suggests that neural activity in separate frequency bands may be responsible for the propagation of predictions and prediction errors respectively \cite{brodski2015,bastos2012}. Future experiments may target information transfer within a specific band to test if the disruption of top-down information transfer happens in the frequency band responsible for the propagation of predictions.

Last, it should be kept in mind that different anesthetics may lead to loss of consciousness by vastly different mechanisms. Ketamine, for example, seems to increase, rather than decrease, overall information transfer---at least in sub-anesthetic doses \cite{rivolta-ketamine-2015}.

\subsection*{Comparison of the alternative approaches to the estimation of information theoretic measures and to statistical testing}

The main analysis of this study was based on information theoretic estimators relying on distances between neighboring data points and  on a permutation  ANOVA. As each of these has its weaknesses we used additional  alternative approaches, first, a Bayesian estimator for information theoretic measures, and second, statistical testing of non-aggregated data using LMMs. Both approaches returned results qualitatively very similar to those of the main analysis. Specifically, replacing neighbor-distance based estimators with Bayesian variants, we replicated the main finding of our study---a reduction in information transfer and locally available information, and an increase in information storage under anesthesia. However, using the Bayesian estimators we did not find a predominant reduction of top-down compared to bottom up information transfer. In contrast, using LMMs for statistical testing instead of a pANOVA additionally revealed a more pronounced reduction in top-down information transfer also in ferret 2 (in this animal, this effect was not significant when performing permutation tests based on the aggregated data).

The Bayesian estimators performed slightly worse than the next neighbor-based estimators in terms of their higher variance in estimates across recordings. Thus, even though Bayesian estimators are currently the best available estimators for discrete data, they may be a non-optimal choice for continuous data. A potential reason for this is the destruction of information on neighborhood relationships through data binning. This is, however, necessary to make the current Bayesian estimators applicable to continuous data.

In sum, we obtained similar results through three different approaches. This makes us confident that locally available information and information transfer indeed decrease under anesthesia, while the amount of predictable information increases.

\subsection*{On information theoretic measures obtained from from continuous time processes via time-discrete sampling}
When interpreting the information theoretic measures presented in this work, it must be kept in mind that they were obtained via time-discrete sampling of processes  that unfold in continuous time (such as the LFP, or spike trains). This time-discrete sampling was taken into account in recent work by Spinney and colleagues \cite{spinney2016}, who could show that the classic transfer entropy as defined by Schreiber is indeed ambiguous when applied to sampled data from time-continuous processes---it can either be seen as the integral of a transfer entropy rate in continuous time over one sampling interval, and is then given in bits, or it can be seen as an approximation to the continuous time transfer entropy rate itself, and be given in bits/sec. In our study we stick with the first notion of transfer entropy, and note that our results will numerically change when using different sampling intervals. In contrast to the case of transfer entropy covered in \cite{spinney2016}, corresponding analytical results for AIS are not known at present, as this is still a field of active research. To nevertheless elucidate the practical dependency of AIS on the sampling rate, when using our estimators, we have taken the original data and have up- and down-sampled them. Indeed the empirical AIS depended on the sampling rate (supporting information S3). Yet, all relations of AIS values between different isoflurane concentrations, i.e., the qualitative results, are independent of sampling. Thus, sampling effects do not affect the conclusions of the current study.

\subsection*{A cautionary note on the interpretation of information theoretic measures evaluated on local field potential data}

When interpreting the results obtained in this study it should be kept in mind that LFP signals are not in themselves the immediate carriers of information relevant to individual neurons. This is because from a neuron's perspective information arrives predominantly in the form postsynaptic potentials generated by incoming spikes and chemical transmission in the synaptic cleft (but see \cite{frohlich2010endogenous} for a potential influence of LFPs on neural dynamics and computation). Thus, LFP signals merely reflect a coarse grained view of the underlying neural information processing. As a consequence, our results only hold in as far as at least some relevant information about the underlying information processing survives this coarse graining in the recording process, and little formal mathematical work has been carried out to estimate bounds on the amount of information available after coarse graining (but see \cite{tao2005szemer}).

Yet, the enormous success that brain reading approaches had when based on local field potentials or on even more coarse grained magnetoencephalography (MEG) recordings (e.g. \cite{roux2012}) indicates that relevant information on neural information processing is indeed available at the level of these signals. However, successful attempts at decoding neural representations of stimuli or other features of the experimental setting should not lead us to misinterpret the information captured by information theoretic measures of neural processing as necessarily \textit{being about something we can understand and link to the outside world}. Quite to the contrary, the larger part of information captured by these measures may be related to intrinsic properties of the unfolding neural computation.

\subsection*{Conclusion}

Using two different methods for transfer entropy estimation, and two different statistical approaches, we found that locally available information and information transfer are reduced under isoflurane administration. The larger decrease in the locally available information was found at the source of the larger decrease of information transfer, not at its end point, or target. Therefore, previously reported reductions in information transfer under anesthesia may be caused by changes in local information processing rather than a disruption of long range connectivity.  We suggest to put this hypothesis more into the focus of future research effort to understand the loss of consciousness under anesthesia. This suggestion receives further support from the fact that the synaptic targets of the anesthetic isoflurane, as used in this study, are most likely located in local circuits.

\section*{Methods}

\subsection*{Electrophysiological Recordings}
\begin{figure}[!h]
    \begin{adjustbox}{center}
        \includegraphics{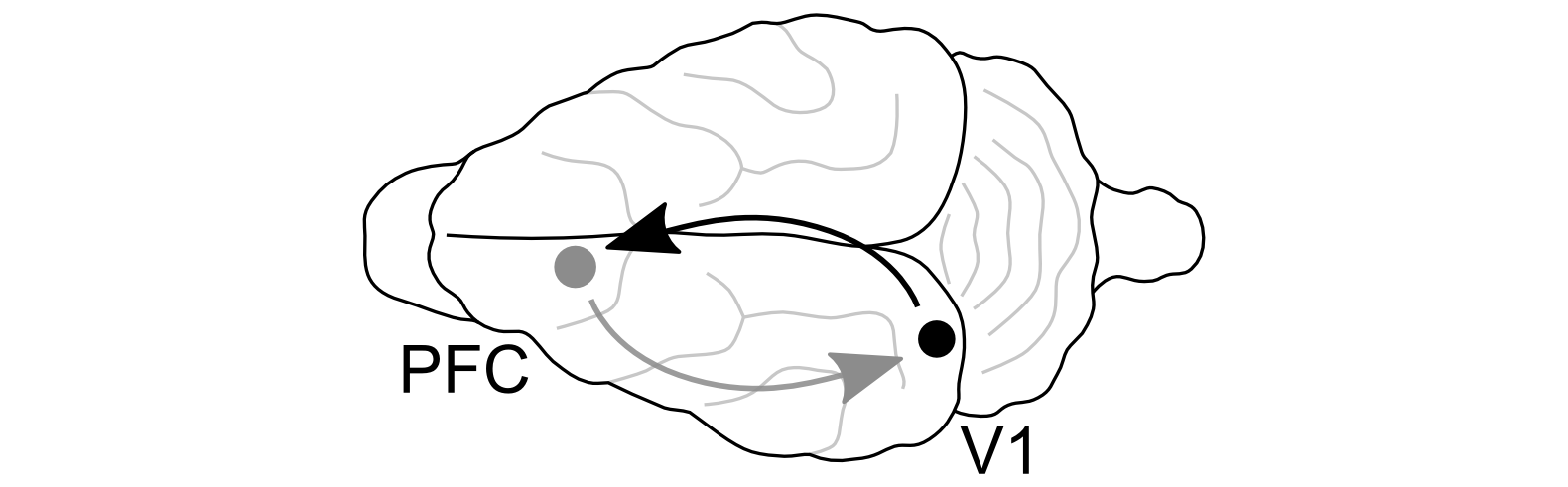}
    \end{adjustbox}
\caption{{\bf Recording sites in the ferret brain.}
Prefrontal cortex (PFC, gray dot): anterior sigmoid gyrus; primary visual area (V1, black dot): lateral gyrus. Arrows indicate analyzed directions of information transfer (gray: top-down; black: bottom-up).}
\label{fig:recording_sites}
\end{figure}

We conducted simultaneous electrophysiological recordings of the local field potential (LFP) in primary visual cortex (V1) and prefrontal cortex (PFC) of two female ferrets (17 to 20 weeks of age at study onset) under different levels of isoflurane (Fig. \ref{fig:recording_sites}). Raw data are available from \cite{raw_data}. The choice of the animal model is discussed further in \cite{sellers2013}. Recordings were made in a dark environment during multiple, individual sessions of max. 2~h length, during which the animals' heads were fixed. For recordings, we used single metal electrodes acutely inserted in putative layer IV, measured 0.3--0.6~mm from the surface of cortex (tungsten micro-electrode, 250-$\mu$m shank diameter, 500-k$\Omega$ impedance, FHC, Bowdoin, ME). The hardware high pass filter was 0.1~Hz and the low pass filter was 5,000~Hz. A silver chloride wire placed between the skull and soft tissue was used as the reference electrode. The reference electrode was located approximately equidistant between the recording electrodes. This location was selected in order to have little shared activity with either recording electrode. The same reference was used for both recording locations; also the same electrode position was used for both animals and all isoflurane concentrations. To verify that electrode placement was indeed in V1, we mapped receptive fields by eliciting visually evoked potentials in a separate series of experiments. We confirmed electrode placement in PFC by lesioning through the recording electrode after completion of data collection and post-mortem histology (as described in \cite{sellers2013}). Details on surgical procedures can be found in \cite{sellers2015}. Unfiltered signals were amplified with gain 1,000 (model 1800, A-M Systems, Carlsborg, WA), digitized at 20~kHz (Power 1401, Cambridge Electronic Design, Cambridge, UK), and digitally stored using Spike2 software (Cambridge Electronic Design). For analysis, data were low pass filtered (300~Hz cutoff) and down-sampled to 1000~Hz.

All procedures were approved by the University of North Carolina-Chapel Hill Institutional Animal Care and Use Committee (UNC-CH IACUC) and exceed guidelines set forth by the National Institutes of Health and U.S. Department of Agriculture.

LFPs were recorded during wakefulness (condition \textit{iso 0.0~\%}, number of recording sessions: 8 and 5 for ferret 1 and 2, respectively) and with different concentrations of anesthetic: 0.5~\% isoflurane with xylazine (condition \textit{iso 0.5~\%}, number of sessions: 5 and 6), as well as 1.0~\% isoflurane with xylazine (condition \textit{iso 1.0~\%}, number of sessions: 10 and 11). In the course of pilot experiments, both concentrations \textit{iso 0.5~\%} and \textit{iso 1.0~\%} lead to a loss of the righting reflex; however, a systematic assessment of this metric during recordings was technically not feasible. Additionally, animals were administered 4.25~ml/h 5~\% dextrose lactated Ringer and 0.015~ml/h xylazine via IV.

LFP recordings from each session were cut into epochs of 4.81~s length to be able to remove segments of data if they were contaminated by artifacts (e.g., due to movement). We chose a relatively short epoch length to avoid removing large chunks of data when there was only a short transient artifact. This resulted in 196 to 513 epochs per recording (mean: 428.6) for ferret 1, and 211 to 526 epochs (mean: 472.8) for ferret 2. epochs with movement artifacts were manually rejected (determined by extreme values in the LFP raw traces). In the \textit{iso 0.0~\%} condition, infrared videography was used to verify that animals were awake during the whole recording; additionally, \textit{iso 0.0~\%} epochs with a relative delta power (0.5 to 4.0~Hz) of more than 30~\% of the total power from 0.5 to 50~Hz were rejected to ensure that only epochs during which the animal was truly awake entered further analysis.

\subsection*{Information theoretic measures}
To measure information transfer between recording sites V1 and PFC, we estimated the transfer entropy \cite{schreiber2000} in both directions of possible interactions, $PFC \rightarrow V1$ and $V1 \rightarrow PFC$. To investigate local information processing within each recording site, we estimated active information storage ($AIS$) \cite{lizier2012LAIS} as a measure of predictable information, and we estimated differential entropy ($H$) \cite{cover2006} as a measure of information available locally. We will now explain the applied measures and estimators in more detail, before we describe how these estimators were applied to data from electrophysiological recordings in the next section. To mathematically formalize the estimation procedure from these data, we assume that neural time series recorded from two systems $\mathcal{X}$ and $\mathcal{Y}$ (e.g. cortical sites) can be treated as collections of realizations $x_t$ and $y_t$ of random variables $X_t$, $Y_t$ of two random processes $X=\{X_1, \ldots , X_t, \ldots, X_N\}$ and $Y=\{Y_1, \ldots , Y_t, \ldots, Y_N\}$. The index $t$ here indicates samples in time, measured in units of the dwell time (inverse sampling rate) of the recording.

\paragraph*{Transfer entropy} Transfer entropy \cite{schreiber2000,vicente2011TE,wibral2014springer} is defined as the mutual information between the future of a process $Y$ and the past of a second process $X$, conditional on the past of $Y$. Transfer entropy thus quantifies the information we obtain about the future of $Y$ from the past of $X$, taking into account information from the past of $Y$. Taking this past of $Y$ into account here removes information redundantly available in the past of both $X$ and $Y$, and reveals information provided synergistically by them \cite{WilliamsII}. In this study, we used an improved estimator of transfer entropy presented in \cite{wibral2013timing}, which accounts for arbitrary information transfer delays:

\begin{align} \label{eq:te}
TE_{SPO}(X \rightarrow Y, t, u) = I \left( Y_t;\mathbf{X}_{t-u}^{d_X}| \mathbf{Y}_{t-1}^{d_Y} \right),
\end{align}

\noindent where $I$ is the conditional mutual information (or the differential conditional mutual information for continuous valued variables) between $Y_t$ and $\mathbf{X}_{t-u}^{d_X}$, conditional on $\mathbf{Y}_{t-1}^{d_Y}$; $Y_t$ is the future value of random process $Y$, and $\mathbf{X}_{t-u}^{d_X}$, $\mathbf{Y}_{t-1}^{d_Y}$ are the past states of $X$ and $Y$, respectively. Past states are collections of past random variables

\begin{align} \label{eq:state}
\mathbf{Y}_{t-1}^{d_Y} = \left( Y_{t-1}, Y_{t-1-\tau}, \ldots, Y_{t-1-(d_Y-1)\tau} \right),
\end{align}

\noindent that form a delay embedding of length $d_Y$ \cite{takens1981}, and that render the future of the random process conditionally independent of all variables of the random process that are further back in time than the variables forming the state. Parameters $\tau$ and $d$ denote the embedding delay and embedding dimension and can be found through optimization of a local predictor as proposed in \cite{ragwitz2002} (see next section on the estimation of information theoretic measures). Past states constructed in this manner are then maximally informative about the present variable of the target process, $Y_t$, which is an important prerequisite for the correct estimation of transfer entropy (see also \cite{wibral2013timing}).

In our estimator (Eq. \ref{eq:te}), the variable $u$ describes the assumed information transfer delay between the processes $X$ and $Y$, which accounts for a physical delay $\delta_{X,Y} \geq 1$ \cite{wibral2013timing}. The estimator thus accommodates arbitrary physical delays between processes. The true delay $\delta_{X,Y}$  must be recovered by 'scanning' various assumed delays and keeping the delay that maximizes $TE_{SPO}$ \cite{wibral2013timing}:

\begin{align} \label{eq:opt_u}
\hat{\delta}_{X,Y} = \argmax_u \left( TE_{SPO}\left( X \rightarrow Y,t,u\right) \right).
\end{align}

\paragraph*{Active Information Storage} $AIS$ \cite{lizier2012LAIS} is defined as the (differential) mutual information between the future of a signal and its immediate past state

\begin{align} \label{eq:ais}
AIS(Y_t) = I \left( Y_t;\mathbf{Y}_{t-1}^{d_Y} \right),
\end{align}

\noindent where $Y$ again is a random process with present value $Y_t$ and past state $\mathbf{Y}_{t-1}^{d_Y}$ (see Eq. \ref{eq:state}). $AIS$ thus quantifies the amount of predictable information in a process or the information that is currently in use for the next state update \cite{lizier2012LAIS}. $AIS$ is low in processes that produce little information or are highly unpredictable, e.g., fully stochastic processes, whereas $AIS$ is highest for processes that visit many equi-probable states in a predictable sequence, i.e., without branching. In other words, $AIS$ is high for processes with ``rich dynamics'' that are predictable from the processes' past \cite{wibral2014LAIS}. A reference implementation of $AIS$ can be found in the Java Information Dynamics Toolkit (JIDT) \cite{lizier2014JIDT}. As for $TE_{SPO}$ estimation, an optimal delay embedding $\mathbf{Y}_{t-1}^{d_Y}$ may be found through optimization of the local predictor proposed in \cite{ragwitz2002}.

 Note, that $AIS$ is upper bounded by the entropy as:

 \begin{align} \label{eq:ais_entropy}
 \begin{split}
 AIS(Y)
 &= I \left( Y_t;\mathbf{Y}_{t-1}^{d_Y} \right)\\
 &= H \left(Y_t\right) - H\left(Y_t|\mathbf{Y}_{t-1}^{d_Y} \right).
 \end{split}
 \end{align}

\paragraph*{Differential entropy} The differential entropy $H$ (see for example \cite{cover2006}) expands the classical concept of Shannon's entropy for discrete variables to continuous variables:

\begin{align} \label{eq:h}
H = - \int_{\mathbb{X}_{t-u}^{d_X}} f(\mathbf{X}_{t-u}^{d_X}) \log f(\mathbf{X}_{t-u}^{d_X}) \, d\mathbf{X}_{t-u}^{d_X},
\end{align}

\noindent where $f(Y_t)$ is the probability density function of $Y_t$ over the support $\mathbb{Y}$. Entropy quantifies the average information contained in a signal. Based on the differential entropy the corresponding measures for mutual and conditional mutual information and, thereby, active information storage and transfer entropy can be defined.

\subsection*{Entropy as an upper bound on information transfer}

The transfer entropy from Eq. \ref{eq:te} can be rewritten as:

\begin{align} \label{eq:bound_conditional_deriv}
\begin{split}
TE_{SPO}(X \rightarrow Y, t, u) =& I \left( \mathbf{X}_{t-u}^{d_X} : Y_t | \mathbf{Y}_{t-1}^{d_Y} \right)\\
=& H( \mathbf{X}_{t-u}^{d_X}|\mathbf{Y}_{t-1}^{d_Y}  ) - H( \mathbf{X}_{t-u}^{d_X}  |\mathbf{Y}_{t-1}^{d_Y}, Y_t  )~,
\end{split}
\end{align}

\noindent By dropping the negative term on the right hand side we obtain an upper bound (as already noticed by ~\cite[p. 65]{faes2014springer}), and by realizing that a conditional entropy is always smaller than the corresponding \textit{unconditional} one, we arrive at

\begin{equation} \label{eq:bound_conditional_1_AIS}
\begin{split}
TE_{SPO}(X \rightarrow Y, t, u) \leq H (\mathbf{X}_{t-u}^{d_X} )
\end{split}
\end{equation}
\noindent This indicates that the overall entropy of the source states is an upper bound. Several interesting other bounds on information transfer exist as detailed in \cite{faes2014springer}, yet these are considerably harder to interpret and were  not the focus of the current presentation.

\subsection*{Estimation of information theoretic measures}

In this section we will describe how the information theoretic measures presented in the last section may be estimated from neural data. In doing so, we will also describe the methodological pitfalls mentioned in the introduction in more detail and we will describe how these were handled here. If not stated otherwise, we used implementations of all presented methods in the open source toolboxes TRENTOOL \cite{lindner2011} and JIDT \cite{lizier2014JIDT}, called through custom MATLAB\textsuperscript{\textregistered{}} scripts (MATLAB 8.0, The MathWorks\textsuperscript{\textregistered{}} Inc., Natick, MA, 2012). Time series were normalized to zero mean and unit variance before estimation.

\paragraph*{Estimating information theoretic measures from continuous data} Estimation of information theoretic measures from continuous data is often handled by simply discretizing the data. This is done either by binning or the use of symbolic time series---mapping the continuous data onto a finite alphabet. Specifically, the use of symbolic time series for transfer entropy estimation was first introduced by \cite{staniek2008STE} and maps the continuous values in past state vectors with length $d$ (Eq. \ref{eq:state}) onto a set of rank vectors. Hence, the continuous-valued time series is mapped onto an alphabet of finite size $d!$. After binning or transformation to rank vectors transfer entropy and the other information theoretic measures can then be estimated using plug-in estimators for discrete data, which simply evaluate the relative frequency of occurrences of symbols in the alphabet. Discretizing the data therefore greatly simplifies the estimation of transfer entropy from neural data, and may even be necessary for very small data sets. Yet, binning ignores the neighborhood relations in the continuous data and the use of symbolic times series destroys important information on the absolute values in the data. An example where transfer entropy estimation fails due to the use of symbolic time series is reported in \cite{pompe2011MIT} and discussed in \cite{wibral2013timing}: In this example, information transfer between two coupled logistic maps was not detected by symbolic transfer entropy \cite{pompe2011MIT}; only when estimating $TE_{SPO}$ directly using an estimator for continuous data, the information transfer was identified correctly \cite{wibral2013timing}. To circumvent the problems with binned or symbolic time series, we here used a nearest-neighbor based $TE_{SPO}$-estimator for continuous data, the Kraskov-St{\"o}gbauer-Grassberger (KSG) estimator for mutual information described in \cite{kraskov2004}. At present, this estimator has the most favorable bias properties compared to similar estimators for continuous data. The KSG-estimator leads to the following expression for the estimation of $TE_{SPO}$ as introduced in Eq. \ref{eq:te} \cite{frenzel2007,vicente2011TE}:

\begin{align} \label{eq:te_est}
\begin{split}
TE_{SPO}(X \rightarrow &Y,t,u) = I(Y_t: \mathbf{X}^{d_X}_{t-u}| \mathbf{Y}^{d_Y}_{t-1}) \\
&= \psi(k) + \langle \psi(n_{\mathbf{y}_{t-1}^{d_Y}} + 1) - \psi(n_{y_t \mathbf{y}_{t-1}^{d_Y}} + 1) - \psi(n_{\mathbf{y}_{t-1}^{d_Y} \mathbf{x}_{t-u}^{d_X}}+1) \rangle_r,
\end{split}
\end{align}

\noindent where $\psi$ denotes the digamma function, $k$ is the number of neighbors  in the highest-dimensional space spanned by variables $Y_{t}$, $\mathbf{Y}^{d_{Y}}_{t-1}$, $\mathbf{X}^{d_{X}}_{t-u}$, and is used to determine search radii for the lower dimensional subspaces; $n_{\cdot}$ are the number of neighbors within these search radii for each point in the lower dimensional search spaces spanned by the variable indicated in the subscript. Angle brackets indicate the average over realizations $r$ (e.g. observations made over an ensemble of copies of the systems or observations made over time in case of stationarity, which we assumed here). We used a $k$ of 4 as recommended by Kraskov \cite[p. 23]{kraskov2004thesis}, such as to balance the estimator's bias---which decreases for larger $k$---and variance---which increases for larger $k$ (see also \cite{khan2007MIest} for similar recommendations based on simulation studies). For a detailed derivation of $TE_{SPO}$-estimation using the KSG-estimator see \cite{kraskov2004,vicente2011TE,frenzel2007}.

The KSG-estimator comes with a bias that is not analytically tractable \cite{kraskov2004thesis}, hence,  estimates can not be interpreted at face value, but have to be tested for their statistical significance against the null-hypothesis of no information transfer \cite{lindner2011,vicente2011TE}. A suitable test distribution for this null-hypothesis can be generated by repeatedly estimating $TE_{SPO}$ from surrogate data. In these surrogate data, the quantity of interest---potential transfer entropy between source and target time series---should be destroyed while all other statistical properties of the data are preserved such that the estimation bias is constant between original and surrogate data. In the present study we thus generated 500 surrogate data sets by permuting whole epochs of the target time series while leaving the order of source time course epochs unchanged (see \cite{wollstadt2014} for a detailed account of surrogate testing).

We used the ensemble-method for transfer entropy estimation \cite{wollstadt2014}, implemented in TRENTOOL \cite{lindner2011}. The ensemble method allows to pool data over epochs, which maximizes the amount of data entering the estimation, while providing an efficient implementation of this estimation procedure using graphics processing units (GPU). We tested the statistical significance of $TE_{SPO}$ in both directions of interaction in the \textit{iso 0.0~\%} condition. We only tested this condition, because $TE_{SPO}$ was expected to be reduced for higher isoflurane levels based on the results of existing studies. Because the estimation of $TE_{SPO}$ is computationally heavy, we used a random subset of 50 epochs to reduce the running time of this statistical test (see supporting information S4 for theoretical and practical running times of the used estimators).
We first tested $TE_{SPO}$ estimates for their significance within individual recording sessions against 500 surrogate data sets, and then used a binomial test to establish the statistical significance \textit{over} recordings. We used a one-sided Binomial test under the null hypothesis of no significant $TE_{SPO}$ estimates, where individual estimates $l$ were assumed to be $B(l,p_0,n)$-distributed, with $p_0=0.05$ and $n=5$ for animal 1 and $p_0=0.05$ and $n=8$ for animal 2.

As the KSG-estimator used for estimating  $TE_{SPO}$ (Eq. \ref{eq:te_est}) is an estimator of mutual information it can also be used for the estimation of $AIS$:

\begin{align} \label{eq:ais_est}
\begin{split}
AIS(Y) &= I(Y_t: \mathbf{Y}^{d_Y}_{t-1}) \\
&= \psi(k) - 1/k + \psi(N) - \langle \psi(n_{y_t}) + \psi(n_{\mathbf{y}^{d_Y}_{t-1}}) \rangle_r,
\end{split}
\end{align}

\noindent where again $\psi$ denotes the digamma function, $k$ is the number of neighbors in the highest-dimensional space, $N$ is the number of realizations, and  $n_{\cdot}$ denotes the number of neighbors for each point in the respective search space. Again, we chose $k=4$ for the estimation of AIS (see above). Note that the sampling rate has an effect on the estimated absolute values of AIS (a simulation of the effect of sampling on AIS estimates are shown as supporting information S3)---however, qualitative results are not influenced by the choice of sampling rate; in other words, relative differences between estimates are the same for different choices of sampling rates. As a consequence, for AIS estimation the sampling rate should be constant over data sets if the aim is to compare these estimates.

A conceptual predecessor of the KSG-estimator for mutual information is the Kozachenko-Leonenko (KL) estimator for differential entropies \cite{kozachenko1987}. The KL-estimator also allows for the estimation of $H$ from continuous data and reads

\begin{align} \label{eq:h_est}
H(X) = -\psi(1) + \psi(N) + \sum_{i=1}^N{\log(\epsilon(i))},
\end{align}

\noindent where $\epsilon(i)$ is twice the distance from data point $i$ to its $k$-th nearest neighbor in the search space spanned by all points.

\paragraph*{Bayesian Estimators for discretized data} For Bayesian estimation we converted the continuous LFP time series to discrete data by applying voltage bins as follows: The voltages $\pm 3$ standard deviations around the mean of the LFP were subdivided into $N$ equally spaced bins. We added two additional bins containing all the values that were either smaller or larger than the 3 SD region, amounting to a total number of bins $N_{bins}=N+2$. We then calculated $H$, $AIS$ and $TE_{SPO}$ for the discrete data. For $AIS$ and $TE_{SPO}$ estimation states were defined using the same dimension $d$ and $\tau$ as for the KSG-estimator, optimized using the Ragwitz criterion. We decomposed $TE_{SPO}$ into four entropies (Eq. \ref{eq:te_est}), and $AIS$ into three entropies, which we then estimated individually \cite{wolpert1995estimating}.  To reduce the bias introduced by the limited number of observed states, we used the NSB-estimator by Nemenman, Shafee, and Bialek \cite{nsb}, which is based on the construction of an almost uniform prior over the expected entropy using a mixture of symmetric Dirichlet priors $\mathcal{P}_{\beta}$. The estimator has been shown to be unbiased for a broad class of distributions that are typical in $\mathcal{P}_{\beta}$ \cite{nsb2}.

We further applied the recently proposed estimator by Archer et al. \cite{archer2014} that uses a prior over distributions with infinite support based on Pitman-Yor-processes. In contrast to the NSB prior, this prior also accounts for heavy-tailed distributions that one encounters frequently in neuronal systems and does not require knowledge of the support of the distribution.

When estimating entropies for the embedding dimensions used here, the number of possible states or ``words'' is between $K=12^{14}$ and $K=12^{31}$. This is much larger than the typical number of observed states per recording of around $N=5\cdot 10^5$. As a consequence, a precise estimation of entropies is only possible if the distribution is sparse, i.e. most words have vanishing probability. In this case, however, the estimates should be independent of the choice of support $K$ as long as $K$ is sufficiently large and does not omit states of finite probability. We chose $K^{\prime}=10^{12}$ for the results shown in this paper, which allowed a robust computation of the NSB estimator instead of the maximum support $K$ that results from simple combinatorics.

\begin{figure}[!h]
   \includegraphics{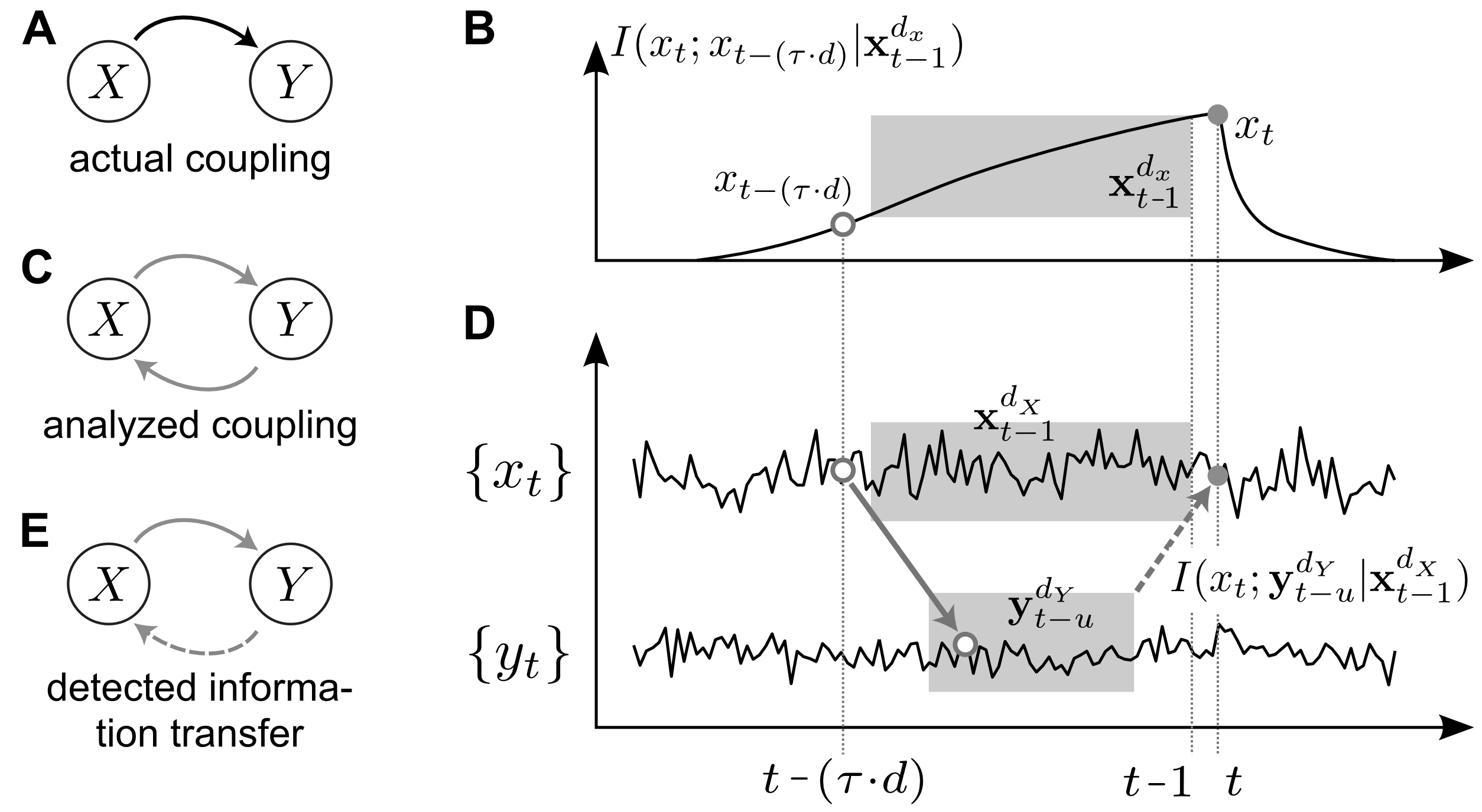}
\caption{{\bf Spurious information transfer resulting from under-embedding (modified from \protect\cite{vicente2011TE}).}
    (A) Actual coupling between processes $X$ and $Y$.
    (B) Mutual information between the present value in $X$, $x_t$, and a value in the far past of $X$, $x_{t-(\tau \cdot d)}$, conditional on all intermediate values $\mathbf{x}_{t-1}^{d_X}$ (shaded box), the mutual information is non-zero, i.e., $x_{t-(\tau \cdot d)}$ holds some information about $x_t$.
    (C) Both directions of interaction are analyzed;
    (D) Information in $x_{t-(\tau \cdot d)}$ (white sample point) is transferred to $Y$ (solid arrow), because of the actual coupling $X \rightarrow Y$. The information in $x_{t-(\tau \cdot d)}$ about $x_t$ is thus transferred to the past of $Y$, $\mathbf{y}_{t-u}^{d_Y}$, which thus becomes predictive of $x_t$ as well. Assume now, we analyzed information transfer from $Y$ to $X$, $I(x_t: \mathbf{y}_{t-u}^{d_Y}|\mathbf{x}_{t-1}^{d_X})$, without a proper embedding of $X$, $\mathbf{x}_{t-1}^{d_X}$: Because of the actually transferred information from $x_{t-(\tau \cdot d)}$ to $\mathbf{y}_{t-u}^{d_Y}$, the mutual information $I(x_t: \mathbf{x}_{t-1}^{d_X})$ is non-zero. If we now under-embed $X$, such that the information in $x_{t-(\tau \cdot d)}$ is not contained in $\mathbf{x}_{t-u}^{d_X}$ and is not conditioned out, $I(x_t:\mathbf{y}_{t-1}^{d_Y}|\mathbf{x}_{t-u}^{d_X})$ will be non-zero as well. In this case, under-embedding of the target $X$ will lead to the detection of spurious information transfer in the non-coupled direction $Y \rightarrow X$.
    (E) Information transfer is falsely detected for both directions of interaction, the link from $Y$ to $X$ is spurious (dashed arrow).}
\label{fig:underembedding}
\end{figure}

\paragraph*{Finding optimal embedding parameters} The second methodological problem raised in the introduction was the choice of embedding parameters for transfer entropy estimation. One important parameter here is the choice of the total signal history when constructing past states for source and target signal (see Eq. \ref{eq:te} and \ref{eq:state}). Failure to properly account for signal histories may lead to a variety of errors, such as underestimating transfer entropy, failure to detect transfer entropy altogether, or the detection of spurious transfer entropy. Transfer entropy is underestimated or missed if the past state of the \textit{source} time series does not cover all the relevant history, i.e., the source is under-embedded. In contrast, spurious transfer entropy may be detected if the past state of the \textit{target} time series is under-embedded, such that spurious detection of transfer entropy is a false positive and therefore the most serious error. One scenario where spurious transfer entropy results from under-embedding is shown in Fig. \ref{fig:underembedding}.

The choice of an optimal embedding is also relevant for the estimation of $AIS$, where under-embedding leads to underestimation of the true $AIS$. Note that on the other hand, we can not increase the embedding length to arbitrarily high values because this leads to computationally intractable problem sizes and requires exponentially more data for estimation.

Optimal embedding parameters $d$ and $\tau$ may be found through the optimization of a local predictor proposed by Ragwitz \cite{ragwitz2002}. Ragwitz' criterion tests different combinations of a range of values for $d$ and $\tau$. The current combination is used to embed each point in a time series, then, the future state for each point is predicted from the future states of its neighbors. The parameter combination that leads to the best prediction on average is used as the optimal embedding. To determine the neighbors of a point, we used a $k$-nearest-neighbor search with $k=4$, i.e., the same value for $k$ as was used for $k$-nearest-neighbor searches when estimating information-theoretic measures. We minimized the mean squared error when optimizing Ragwitz' local predictor. Further details on Ragwitz' criterion can be found in the documentations of the TRENTOOL \cite{lindner2011} and JIDT \cite{lizier2014JIDT} toolbox.

Other approaches for embedding parameter optimization have been proposed, see for example non-uniform embedding using mutual information to determine all relevant past samples as proposed by \cite{faes2011nonuniform}.

\paragraph*{Reconstruction of information transfer delays} The third methodological problem raised in the introduction was failure to account for a physical delay $\delta$ between neural sites when estimating transfer entropy. In our estimator $TE_{SPO}$ (Eq. \ref{eq:te_est}) we account for $\delta$ by introducing the parameter $u$. The delay $u$ needs to be optimized to correctly estimate $TE_{SPO}$. If $u$ is not optimal, i.e., $u$ is not sufficiently close to $\delta$ (Fig. \ref{fig:delay_reconstruction} and \cite{wibral2013timing}), information transfer may be underestimated or not measured at all. This is because choosing the parameter $u$ too large ($u \gg \delta$) means that the information present in the evaluated samples of the source is also present in the history of the target already, and conditioned away. In contrast, choosing the parameter $u$ too small means that the information of the evaluated samples of the source will only arrive in the future of the current target sample, and is useless for providing information about it (Fig. \ref{fig:delay_reconstruction}A).

It can be proven for bivariate systems that $TE_{SPO}$ becomes maximal when the true delay $\delta$ is chosen for $u$ \cite{wibral2013timing}. Therefore, the true delay $\delta$---and thus an optimal choice for $u$---can be found by using the value for $u$ that maximizes $TE_{SPO}$ \cite{wibral2013timing}. This optimal $u$ can be found by scanning a range of assumed values. In the present study, we scanned values ranging from 0 to 20 ms. (Note that assumed values should be physiologically plausible to keep the computations practically feasible.)

Accounting for the information transfer $\delta$ by finding optimal parameters $u$ for $TE_{SPO}$ estimation has important consequences when calculating indices from estimated $TE_{SPO}$, such as $TE_{net}$:

\begin{align}
TE_{net}=\frac{TE_{SPO}(X \rightarrow Y,t,u_{X \rightarrow Y}) - TE_{SPO}(Y \rightarrow X,t,u_{Y \rightarrow X})}{TE_{SPO}(X \rightarrow Y,t,u_{X \rightarrow Y}) + TE_{SPO}(Y \rightarrow X,t,u_{Y \rightarrow X})},
\label{eq:te_net}
\end{align}

\noindent or variations of this measure. The $TE_{net}$ is popular in anesthesia research \cite{lee2013,untergehrer2014} and indicates the predominant direction of information transfer between two bidirectionally coupled processes $X$ and $Y$ ($TE_{net} \geq 0$ if $TE_{SPO}(X \rightarrow Y,t,u_{X \rightarrow Y}) \geq TE_{SPO}(Y \rightarrow X,t,u_{Y \rightarrow X})$ and $TE_{net} < 0$ if $TE_{SPO}(Y \rightarrow X,t,u_{Y \rightarrow X}) > TE_{SPO}(X \rightarrow Y,t,u_{X \rightarrow Y})$). However, if values for $u_{X \rightarrow Y}$ and $u_{Y \rightarrow X}$ are not optimized individually, $TE_{net}$ may take on arbitrary signs: In Fig. \ref{fig:delay_reconstruction}B, we show a toy example of two coupled Lorenz systems \cite{wibral2013timing}, where the absolute difference between raw $TE_{SPO}$ values changes as a function of a common $u$ for both directions and where the difference even changes signs for values $u>65$. To obtain a meaningful value from $TE_{net}$ we thus need to find the individually optimal choices of $u$ for both directions of transfer---in the example, these optima are found at $u_{X\rightarrow Y} = 46$ and $u_{Y \rightarrow X} = 76$, leading to the ``true'' difference.

\paragraph*{Simulating the effect of filtering on information transfer delay reconstruction} To simulate the effects of filtering as a preprocessing technique on the ability of the $TE_{SPO}$ estimator to reconstruct the correct information transfer delay, we simulated two coupled time series, for which we estimated transfer entropy before filtering and after band-pass filtering with different bandwidths. The simulation was repeated 50 times.

We simulated two time series with 100,000 samples each, which were drawn from a uniform random distribution over the open interval (0,1). We introduced a coupling between the time series by adding a scaled version (factor 0.2) of the first time series to the second with a delay of 10~samples. We estimated $TE_{SPO}$ from the first to the second time series using the KSG-estimator implemented in the JIDT toolbox, with $k=4$ and a history of one sample for both source and target.

We estimated $TE_{SPO}$ with and without band-pass filtering the data for different values of $u$, ranging from 1 to 20~samples. We filtered the data using a fourth order, causal Butterworth filter, implemented in the MATLAB toolbox FieldTrip \protect\cite{oostenveld2011FT}. We filtered the data using four different bandwidths: from 0.1 to 300~Hz (corresponding to the filtering done in this study), from 0.1 to 200~Hz, from 12 to 30~Hz (corresponding to the beta frequency range), and from 4 to 8~Hz (corresponding to the theta frequency range).

\subsection*{Statistical testing using permutation testing}

To test for statistically relevant effects of isoflurane levels and direction of interaction or recording site on estimated measures, we performed two-factorial permutation analyses of variance (pANOVA) for each animal and estimated measure \cite{anderson2003pANOVA,suckling2004pANOVA,helbling2015,brodski2015}. We used a MATLAB\textsuperscript{\textregistered{}} implementation of the test described in \cite{helbling2015}, which is compatible with the FieldTrip toolbox data format \cite{oostenveld2011FT} and is available from \cite{helbling2017github}.

The permutation ANOVA can be used if the normality assumptions of parametric ANOVA are violated or---as in the present study---if assumptions are not testable due to too few data points per factor level. The permutation ANOVA evaluates the significance of the main effect of individual factors or their interaction effect under the null-hypothesis of no experimental effect at all. The significance is evaluated by calculating a F-ratio for the effect from the original data; this original F-ratio is then compared against a distribution of F-ratios obtained from permuted data. The F-ratio's p-value is calculated as the fraction of ratios obtained from permuted data that is bigger than the original F-ratio. When permuting data, it is crucial to only permute data in such a way that the currently investigated effect is destroyed while all other effects are kept intact \cite{anderson2003pANOVA}: for example, consider a two-factorial design with factors $A$ and $B$---if the main effect of factor $A$ is to be tested, the assignment of levels of $A$ to data points has to be permuted; yet, the permutation of levels of $A$ can only happen \textit{within} levels of factor $B$ such as to not simultaneously destroy the effect of factor $B$. Thus, when testing for the effect of one factor, the effect of all other factors is preserved, making sure that variability due to one factor is tested for while the variability due to the other factor is held constant. However, this permutation scheme is not applicable to the interaction effect because it leaves no possible permutations---destroying the interaction effect through permutation always also destroys the main effects of individual factors. Here, both factors have to be permuted, which yields an approximative test of the interaction effect (see \cite{helbling2015} for a discussion of permutation strategies and simulations).

The factors in the permutation ANOVA were \textit{isoflurane level} and \textit{direction} for $TE_{SPO}$ estimates, and \textit{isoflurane level} and \textit{recording site} for $AIS$ and $H$ estimates. The number of permutations was set to 10,000. In the present study, data was recorded in different sessions and segmented into epochs. As a result, estimates for individual epochs should not be pooled over recordings for statistical analysis (see for example \cite{aarts2014}, and also the next paragraph for a discussion). We therefore aggregated estimates over epochs to obtain one estimate per recording session. We used the median to aggregate the estimated values for each recording session over individual epochs, because the distribution of measures over epochs was skewed and we considered the median a more exact representation of the distributions' central tendencies. The aggregation of data resulted in relatively few observations per ANOVA cell and also in unequal number of observations between cells, thus violating two basic assumptions of parametric ANOVA (too few observations make it impossible to test for parametric assumptions like homogeneity of variances). Therefore, we used the non-parametric permutation approach over a parametric one, because the permutation ANOVA does not make any assumptions on data structure.

\subsection*{Statistical testing using linear mixed models} \label{lbl:suppl_lmm}
As described above, LFP recordings were conducted in epochs over multiple recording sessions. This introduces a so-called ``nested design'' \cite{aarts2014}, i.e., a hierarchical structure in the data, where data epochs are nested within recordings, which are nested in animals. Such structures lead to systematic errors or dependence within the data. This violates the assumption of uncorrelated errors made by the most common tests derived from the general linear model and leads to an inflation of the type I error \cite{aarts2014}. A measure of the degree of dependence is the intraclass correlation coefficient (ICC), which was 0.35 for ferret 1 and 0.19 for ferret 2, indicating a significant dependency within the data (see \cite{aarts2014} for a discussion of this measure). Thus, we performed an additional statistical tests where we again tested for a significant effect of the two factors \textit{isoflurane level} and \textit{direction} as well as their interaction, but we additionally modeled the nested structure in our data by a random factor \textit{recordings}. Such a model is called a linear mixed effects model \cite{fahrmeir2007regression}, and may yield higher statistical power than aggregating data within one level of the nested design (as was done for the permutation ANOVA). We used the following model for both animals separately:

$$
\label{eq:lmm_model}
y_{ij} = \beta_0 + \gamma_{j} + \beta_1 D_i + \beta_2 A^{(0.5)}_i + \beta_3 A^{(1.0)}_i + \epsilon_{ij},
$$

where $y_{ij}$ is the $TE_{SPO}$ value from the $i$-th epoch in recording $j$, modeled as a function of isoflurane level and direction of interaction, where $\beta_0$ describes the model intercept, $\beta_{k}$ are the regression coefficients and describe fixed effects, $\gamma_j$ is the random deviation of recording $j$ from the intercept $\beta_0$, and $\epsilon_{ij}$ describes random noise. $D_i$, $A^{(0.5)}_i$, and $A^{(1.0)}_i$ are predictor variables, encoding factors \textit{direction} as

\begin{equation}
D_i :=
\begin{cases}
1 & \textrm{if } \textrm{ \textit{direction} is } PFC \rightarrow V1, \\
-1 & \textrm{if } \textrm{ \textit{direction} is } V1 \rightarrow PFC,
\end{cases}
\label{eq:cases_factor_direction}
\end{equation}

\noindent and factor \textit{isoflurane level} as

\begin{equation}
A^{(0.5)}_i :=
\begin{cases}
1 & \textrm{if \textit{isoflurane level} is \textit{iso 0.5 \%},} \\
0 & \textrm{else,}
\end{cases}
\label{eq:cases_factor_anesthesia_1}
\end{equation}

\begin{equation}
A^{(1.0)}_i :=
\begin{cases}
1 & \textrm{if \textit{isoflurane level} is \textit{iso 1.0 \%},} \\
0 & \textrm{else.}
\end{cases}
\label{eq:cases_factor_anesthesia_2}
\end{equation}

\noindent Note that we used dummy coding for factor \textit{direction} so that the the estimated effect $\beta_1$ can be interpreted like the simple or main effect in a standard ANOVA framework. We used contrast coding for factor \textit{isoflurane level} which allows to interpret estimated effects $\beta_2$ and $\beta_3$ as deviations from a reference group (in this case the condition \textit{iso 0.0~\%}). We further assume that noise was i.i.d. and $\epsilon_{ij} \sim \mathcal{N}\left(0, \sigma^2_{\epsilon}\right)$ and $\gamma_j \sim \mathcal{N}\left(0, \sigma^2_{\gamma}\right)$.

We used the R language \cite{Rlanguage} and the function \texttt{lmer} from the \texttt{lme4}-package \cite{lme4} for model fitting. We assessed statistical significance of individual factors by means of model comparison using the maximum likelihood ratio between models \cite{barr2013}. To allow for this model comparison, we used maximum likelihood estimation, instead of restricted maximum likelihood estimation, of random and fixed effects. 
To test for main effects, we compared models including individual factors \textit{isoflurane level} ($fm\_a$) and \textit{direction} ($fm\_d$) to a Null model ($fm\_0$) including only the random effect; to furthermore test for an interaction effect, we compared the model including an interaction term ($fm\_axb$) to a model where both factors only entered additively ($fm\_ab$).

The models were fitted to 15,973 $TE_{SPO}$ values from ferret 1 and 18,202 $TE_{SPO}$ values from ferret 2, respectively.

\subsection*{Simulating the effect of reduced source entropy on transfer entropy}

To test whether changes in source entropy influenced the transfer entropy despite unchanged coupling, we simulated two test cases, with high and low source entropy, respectively, while the coupling between source and target process were held constant. To simulate the two test cases, we randomly selected two recordings---one from the \textit{iso 0.0~\%} condition, which on average showed higher source entropy, and one from the \textit{iso 1.0~\%} condition, which on average showed lower source entropy (Figs. \ref{fig:pANOVA_main_effects_all} and \ref{fig:pANOVA_main_effects_bayes}). In the recording from the \textit{iso 0.0~\%} condition, we permuted epochs in the target time series to destroy all information transfer present in the original data. From the permuted data, we simulated two cases of artificial coupling, first, using the high-entropy source time course from the \textit{iso 0.0~\%} condition; and second, using the low-entropy source time course from the \textit{iso 1.0~\%} condition. The coupling was simulated by adding a filtered, scaled and delayed version of the respective source time course to the target time course for each epoch. For filtering, we used a Gaussian filter with a smoothing of 10 samples; for the scaling factor, we used a value of 0.2, which resulted in a $TE_{SPO}$ value for the high-entropy test case that was close to the $TE_{SPO}$ in the \textit{iso 0.0~\%} condition. By replacing the original coupling in both recordings with a simulated coupling, we made sure that the coupling was constant for both test cases.

For both test cases, we estimated $TE_{SPO}$ and $H(\mathbf{X}_{t-u}^{d_X})$  following the estimation procedures described above. We tested differences in $TE_{SPO}$ using a permutation independent samples t-test with 10,000 permutations. To make sure that our simulation reflected information transfer found in the original data, we further tested for a significant difference between $TE_{SPO}$ in the original data from the \textit{iso 0.0~\%} recording and $TE_{SPO}$ in the high-entropy test case (using the source time \textit{iso 0.0~\%} condition). Here, we found no significant difference, indicating that the transfer entropy in our test case did not differ significantly from the transfer entropy found in the original data.

\subsection*{Correlating information-theoretic measures with other time-series properties} We correlated $TE_{SPO}$, $AIS$, and $H$ with more conventional measures from time-series analysis, namely, the autocorrelation decay time (ACT), signal variance, and power in individual frequency bands. This correlation was performed to investigate whether information-theoretic measures captured signal properties that could be described equally well by more simple measures.

The ACT was calculated by finding the lag at which the autocorrelation coefficient decayed below $e^{-1}$. The signal variance was calculated as the variance over time for each recording after subtracting the mean of the time series. The band power was calculated for individual frequency bands (delta = 0.5--4~Hz, theta = 4--8~Hz, alpha = 8--12~Hz, beta = 12--30~Hz, gamma = 30--40~Hz, following \cite{sellers2015}), using the multitaper method with discrete prolate spheroidal sequences (Slepian sequences) as tapers (smoothing = 1~Hz) implemented in the MATLAB toolbox FieldTrip \cite{oostenveld2011FT}.

We calculated correlations between information-theoretic measures and conventional measures for individual isoflurane levels and recording sites or direction of interaction, respectively. For each correlation, we pooled data over recording sessions for the respective isoflurane level and calculated Spearman's rank correlation. We tested the correlation for significance using a restricted permutation test, where permutations were allowed only \textit{within} one recording, accounting for the nested experimental design (see \cite{aarts2014} and section \textit{Statistical testing using linear mixed models}, below). Additionally, we calculated the correlation as well as the variance explained, $R^2$, for individual recordings, because calculating $R^2$ for the coefficient calculated from \textit{pooled} data does not yield interpretable results.

\section*{Acknowledgements} This study was partly inspired by the work of Vasily A.\,Vakorin on the interplay of complexity and information transfer in \cite{vakorin2014springer}. The authors thank Saskia Helbling for fruitful discussions on data analysis and for making the permutation ANOVA code available. The authors acknowledge support by the Human Brain Project (EU Grant No. 604102).

\section*{Supporting Information}

\paragraph*{S1 Dataset}
\label{S1_data}
{\bf Group statistical data and estimated information-theoretic values entering statistical tests.}

\paragraph*{S2 Correlation tables}
\label{S2_corr_tables}
{\bf Correlation between information-theoretic measures and traditional measures of time-series properties.}

\begin{table}[H]
\caption{Correlation of $AIS$ with autocorrelation decay time (ACT). $\widetilde{R}^2$ and $max(R^2)$ indicate the median and maximum of $R^2$ over recordings per condition, respectively.}
\label{tab:ais_act}
\begin{tabular}{lllcc}
\hline
\textbf{animal} & \multicolumn{1}{p{1.7cm}}{\textbf{isoflurane level}} & \multicolumn{1}{p{1.7cm}}{\textbf{recording site}} & $p$ & $\widetilde{R}^2$ ($max(R^2)$ \\ \hline
1 & iso 0.0 \% & PFC & 1.0000         & 0.024 (0.289) \\
  &            & V1  & 0.1028         & 0.003 (0.153) \\
  & iso 0.5 \% & PFC & 0.0013$^{**}$  & 0.001 (0.122) \\
  &            & V1  & 0.0000$^{***}$ & 0.014 (0.114) \\
  & iso 1.0 \% & PFC & 0.0000$^{***}$ & 0.005 (0.271) \\
  &            & V1  & 0.9998         & 0.004 (0.051) \\ \hline

2 & iso 0.0 \% & PFC & 0.0000$^{***}$ & 0.022 (0.116) \\
  &            & V1  & 0.0000$^{***}$ & 0.029 (0.175) \\
  & iso 0.5 \% & PFC & 0.7672         & 0.007 (0.018) \\
  &            & V1  & 0.0000$^{***}$ & 0.017 (0.076) \\
  & iso 1.0 \% & PFC & 0.0000$^{***}$ & 0.008 (0.121) \\
  &            & V1  & 0.8247         & 0.009 (0.214) \\ \hline
\multicolumn{5}{l}{$^* p<0.05$; $^{**} p<0.01$; $^{***} p<0.001$} \\
\end{tabular}
\vspace*{-4pt}
\end{table}

\begin{table}[H]
\caption{Correlation of $H$ with signal variance. $\widetilde{R}^2$ and $max(R^2)$ indicate the median and maximum of $R^2$ over recordings per condition, respectively.}
\label{tab:h_signal_var}
\begin{tabular}{lllcc}
\hline
\textbf{animal} & \multicolumn{1}{p{1.7cm}}{\textbf{isoflurane level}} & \multicolumn{1}{p{1.7cm}}{\textbf{recording site}} & $p$ & $\widetilde{R}^2$ ($max(R^2)$ \\ \hline
1 & iso 0.0 \% & PFC & 1.0000 & 0.087 (0.705) \\
  &            & V1  & 1.0000 & 0.079 (0.452) \\
  & iso 0.5 \% & PFC & 1.0000 & 0.036 (0.093) \\
  &            & V1  & 1.0000 & 0.062 (0.217) \\
  & iso 1.0 \% & PFC & 1.0000 & 0.010 (0.474) \\
  &            & V1  & 1.0000 & 0.018 (0.083) \\ \hline

2 & iso 0.0 \% & PFC & 1.0000 & 0.042 (0.699) \\
  &            & V1  & 1.0000 & 0.162 (0.363) \\
  & iso 0.5 \% & PFC & 1.0000 & 0.041 (0.130) \\
  &            & V1  & 1.0000 & 0.031 (0.289) \\
  & iso 1.0 \% & PFC & 1.0000 & 0.118 (0.602) \\
  &            & V1  & 1.0000 & 0.018 (0.578) \\ \hline
\multicolumn{5}{l}{$^* p<0.05$; $^{**} p<0.01$; $^{***} p<0.001$} \\
\end{tabular}
\vspace*{-4pt}
\end{table}

\begin{table}[H]
\caption{Correlation of $TE_{SPO}$ with source autocorrelation decay time (ACT) and source signal variance. $\widetilde{R}^2$ and $max(R^2)$ indicate the median and maximum of $R^2$ over recordings per condition, respectively.}
\label{tab:te_act_signal_var}
\begin{adjustbox}{center}
\begin{tabular}{lllcccc}
\hline
\textbf{animal} & \multicolumn{1}{p{1.7cm}}{\textbf{isoflurane level}} & \multicolumn{1}{p{1.7cm}}{\textbf{direction}} & \multicolumn{2}{c}{ACT} & \multicolumn{2}{c}{signal variance}\\
& & & $p$ & $\widetilde{R}^2$ ($max(R^2)$ & $p$ & $\widetilde{R}^2$ ($max(R^2)$ \\ \hline
1 & iso 0.0 \% & PFC $\rightarrow$ V1 & 1.0000       & 0.035 (0.110)  & 0.9861         & 0.009 (0.060) \\
  &            & V1 $\rightarrow$ PFC & 0.9996       & 0.029 (0.215)  & 0.6644         & 0.024 (0.089) \\
  & iso 0.5 \% & PFC $\rightarrow$ V1 & 1.0000       & 0.016 (0.110)  & 1.0000         & 0.024 (0.033) \\
  &            & V1 $\rightarrow$ PFC & 0.7310       & 0.003 (0.014)  & 0.0063$^{*}$   & 0.005 (0.011) \\
  & iso 1.0 \% & PFC $\rightarrow$ V1 & 1.0000       & 0.006 (0.047)  & 0.5836         & 0.002 (0.037) \\
  &            & V1 $\rightarrow$ PFC & 0.0069$^{*}$ & 0.001 (0.032)  & 0.0000$^{***}$ & 0.005 (0.092) \\ \hline
2 & iso 0.0 \% & PFC $\rightarrow$ V1 & 1.0000        & 0.063 (0.105) & 0.8315         & 0.008 (0.025)  \\
  &            & V1 $\rightarrow$ PFC & 0.9996        & 0.006 (0.041) & 0.0434         & 0.005 (0.031)  \\
  & iso 0.5 \% & PFC $\rightarrow$ V1 & 1.0000        & 0.115 (0.155) & 0.0440         & 0.017 (0.103)  \\
  &            & V1 $\rightarrow$ PFC & 1.0000        & 0.022 (0.125) & 0.9985         & 0.005 (0.086)  \\
  & iso 1.0 \% & PFC $\rightarrow$ V1 & 0.0000$^{**}$ & 0.011 (0.111) & 0.0000$^{***}$ & 0.012 (0.058)  \\
  &            & V1 $\rightarrow$ PFC & 0.0076$^{*}$  & 0.009 (0.031) & 0.0000$^{***}$ & 0.015 (0.032)  \\ \hline
\multicolumn{7}{l}{$^* p<0.05$; $^{**} p<0.01$; $^{***} p<0.001$} \\
\end{tabular}
\end{adjustbox}
\vspace*{-4pt}
\end{table}

\begin{table}[H]
\caption{Correlation of $AIS$ with band power. $\widetilde{R}^2$ and $max(R^2)$ indicate the median and maximum of $R^2$ over recordings per condition, respectively.}
\label{tab:ais_band_pow}
\begin{adjustbox}{center}
\begin{tabular}{lllcccccc}
\hline
\textbf{animal} & \textbf{level} & \textbf{site} & \multicolumn{2}{c}{$\delta$} & \multicolumn{2}{c}{$\theta$} & \multicolumn{2}{c}{$\alpha$}\\
& & & $p$ & $\widetilde{R}^2$ ($max(R^2)$ & $p$ & $\widetilde{R}^2$ ($max(R^2)$ & $p$ & $\widetilde{R}^2$ ($max(R^2)$ \\ \hline
1 & iso 0.0 \% & PFC & 1.0000         & 0.068 (0.418) & 0.0000$^{***}$ & 0.021 (0.201) & 0.0000$^{***}$ & 0.022 (0.258) \\
  &            & V1  & 0.0000$^{***}$ & 0.041 (0.275) & 0.0003$^{**}$  & 0.020 (0.160) & 0.0000$^{***}$ & 0.029 (0.166) \\
  & iso 0.5 \% & PFC & 0.0000$^{***}$ & 0.019 (0.174) & 0.0000$^{***}$ & 0.005 (0.040) & 0.0000$^{***}$ & 0.003 (0.075) \\
  &            & V1  & 0.0000$^{***}$ & 0.008 (0.129) & 0.0093         & 0.024 (0.060) & 0.0986         & 0.009 (0.090) \\
  & iso 1.0 \% & PFC & 0.0000$^{***}$ & 0.006 (0.036) & 0.8710         & 0.005 (0.093) & 0.0445         & 0.003 (0.028) \\
  &            & V1  & 0.2751         & 0.010 (0.303) & 0.0000$^{***}$ & 0.002 (0.054) & 0.0000$^{***}$ & 0.002 (0.030) \\ \hline

2 & iso 0.0 \% & PFC & 0.0000$^{***}$ & 0.158 (0.359) & 0.0000$^{***}$ & 0.044 (0.089) & 0.0000$^{***}$ & 0.025 (0.075) \\
  &            & V1  & 0.0000$^{***}$ & 0.005 (0.303) & 0.0000$^{***}$ & 0.023 (0.228) & 0.0000$^{***}$ & 0.042 (0.238) \\
  & iso 0.5 \% & PFC & 0.0000$^{***}$ & 0.019 (0.255) & 0.0000$^{***}$ & 0.007 (0.018) & 0.0000$^{***}$ & 0.006 (0.074) \\
  &            & V1  & 0.0000$^{***}$ & 0.025 (0.098) & 0.6476         & 0.024 (0.068) & 1.0000         & 0.017 (0.045) \\
  & iso 1.0 \% & PFC & 0.0000$^{***}$ & 0.005 (0.298) & 0.9729         & 0.005 (0.021) & 0.9999         & 0.004 (0.151) \\
  &            & V1  & 0.0063$^{*}$   & 0.043 (0.302) & 0.7462         & 0.001 (0.039) & 1.0000         & 0.003 (0.051) \\
&&&&&&&& \\
&&&&&&&& \\ \hline
\textbf{animal} & \textbf{level} & \textbf{site}  & \multicolumn{2}{c}{$\beta$}  & \multicolumn{2}{c}{$\gamma$} & \\
&        &     & $p$ & $\widetilde{R}^2$ ($max(R^2$) & $p$ & $\widetilde{R}^2$ ($max(R^2)$ & \\ \hline
1 & iso 0.0 \% & PFC & 0.1179         & 0.016 (0.154) & 1.0000 & 0.036 (0.089) & \\
  &            & V1  & 0.0001$^{***}$ & 0.039 (0.163) & 1.0000 & 0.018 (0.331) & \\
  & iso 0.5 \% & PFC & 0.0000$^{***}$ & 0.003 (0.023) & 0.5147 & 0.007 (0.068) & \\
  &            & V1  & 0.9981         & 0.025 (0.081) & 1.0000 & 0.005 (0.009) & \\
  & iso 1.0 \% & PFC & 0.7781         & 0.001 (0.044) & 0.9392 & 0.004 (0.063) & \\
  &            & V1  & 0.1122         & 0.003 (0.045) & 0.8076 & 0.003 (0.061) & \\ \hline

2 & iso 0.0 \% & PFC & 0.0000$^{***}$ & 0.005 (0.047)  & 1.0000         & 0.006 (0.077) & \\
  &            & V1  & 0.2450         & 0.015 (0.185)  & 1.0000         & 0.027 (0.060) & \\
  & iso 0.5 \% & PFC & 0.0000$^{***}$ & 0.007 (0.094)  & 0.0000$^{***}$ & 0.014 (0.088) & \\
  &            & V1  & 1.0000         & 0.036 (0.087)  & 1.0000         & 0.010 (0.080) & \\
  & iso 1.0 \% & PFC & 1.0000         & 0.007 (0.073)  & 1.0000         & 0.008 (0.227) & \\
  &            & V1  & 1.0000         & 0.001 (0.062)  & 1.0000         & 0.017 (0.147) & \\ \hline

\multicolumn{7}{l}{$^* p<0.05$; $^{**} p<0.01$; $^{***} p<0.001$} \\
\end{tabular}
\end{adjustbox}
\vspace*{-4pt}
\end{table}

\begin{table}[H]
\caption{Correlation of $H$ with band power. $\widetilde{R}^2$ and $max(R^2)$ indicate the median and maximum of $R^2$ over recordings per condition, respectively.}
\label{tab:h_band_pow}
\begin{adjustbox}{center}
\begin{tabular}{lllcccccc}
\hline
\textbf{animal} & \textbf{level} & \textbf{site} & \multicolumn{2}{c}{$\delta$} & \multicolumn{2}{c}{$\theta$} & \multicolumn{2}{c}{$\alpha$}\\
& & & $p$ & $\widetilde{R}^2$ ($max(R^2)$ & $p$ & $\widetilde{R}^2$ ($max(R^2)$ & $p$ & $\widetilde{R}^2$ ($max(R^2)$ \\ \hline
1 & iso 0.0 \% & PFC & 1.0000 & 0.095 (0.458) & 1.0000         & 0.015 (0.201)  & 1.0000         & 0.020 (0.271) \\
  &            & V1  & 1.0000 & 0.091 (0.706) & 0.9998         & 0.037 (0.109)  & 1.0000         & 0.041 (0.097) \\
  & iso 0.5 \% & PFC & 1.0000 & 0.046 (0.202) & 1.0000         & 0.004 (0.036)  & 1.0000         & 0.003 (0.059) \\
  &            & V1  & 1.0000 & 0.035 (0.133) & 0.9100         & 0.040 (0.051)  & 0.8029         & 0.002 (0.086) \\
  & iso 1.0 \% & PFC & 1.0000 & 0.020 (0.079) & 0.0359         & 0.005 (0.082)  & 0.9198         & 0.004 (0.023) \\
  &            & V1  & 1.0000 & 0.010 (0.461) & 0.9941         & 0.003 (0.061)  & 0.9991         & 0.004 (0.032) \\ \hline
2 & iso 0.0 \% & PFC & 1.0000 & 0.171 (0.395) & 1.0000         & 0.058 (0.110)  & 1.0000         & 0.024 (0.060) \\
  &            & V1  & 1.0000 & 0.031 (0.723) & 1.0000         & 0.036 (0.135)  & 1.0000         & 0.048 (0.170) \\
  & iso 0.5 \% & PFC & 1.0000 & 0.035 (0.298) & 1.0000         & 0.011 (0.016)  & 1.0000         & 0.015 (0.088) \\
  &            & V1  & 1.0000 & 0.038 (0.135) & 0.0428         & 0.021 (0.073)  & 0.0000$^{***}$ & 0.016 (0.045) \\
  & iso 1.0 \% & PFC & 1.0000 & 0.015 (0.580) & 0.0000$^{***}$ & 0.006 (0.076)  & 0.0000$^{***}$ & 0.008 (0.219) \\
  &            & V1  & 1.0000 & 0.112 (0.601) & 0.1329         & 0.003 (0.062)  & 0.0000$^{***}$ & 0.009 (0.094) \\

&&&&&&&& \\
&&&&&&&& \\ \hline
\textbf{animal} & \textbf{level} & \textbf{site}  & \multicolumn{2}{c}{$\beta$}  & \multicolumn{2}{c}{$\gamma$} & \\
&        &     & $p$ & $\widetilde{R}^2$ ($max(R^2$) & $p$ & $\widetilde{R}^2$ ($max(R^2)$ & \\ \hline
1 & iso 0.0 \% & PFC & 0.9665         & 0.018 (0.165) & 0.0000$^{***}$ & 0.052 (0.097) & \\
  &            & V1  & 0.9999         & 0.046 (0.121) & 0.0000$^{***}$ & 0.014 (0.358) & \\
  & iso 0.5 \% & PFC & 1.0000         & 0.007 (0.020) & 0.5510         & 0.013 (0.064) & \\
  &            & V1  & 0.0000$^{***}$ & 0.028 (0.089) & 0.0000$^{***}$ & 0.006 (0.011) & \\
  & iso 1.0 \% & PFC & 0.2144         & 0.005 (0.056) & 0.0269         & 0.006 (0.082) & \\
  &            & V1  & 0.4394         & 0.002 (0.040) & 0.2250         & 0.003 (0.053) & \\ \hline
2 & iso 0.0 \% & PFC & 1.0000         & 0.011 (0.043) & 0.0000$^{***}$ & 0.001 (0.068) & \\
  &            & V1  & 0.5631         & 0.022 (0.094) & 0.0000$^{***}$ & 0.037 (0.061) & \\
  & iso 0.5 \% & PFC & 1.0000         & 0.013 (0.107) & 1.0000         & 0.012 (0.095) & \\
  &            & V1  & 0.0000$^{***}$ & 0.034 (0.096) & 0.0000$^{***}$ & 0.010 (0.102) & \\
  & iso 1.0 \% & PFC & 0.0000$^{***}$ & 0.007 (0.051) & 0.0000$^{***}$ & 0.012 (0.290) & \\
  &            & V1  & 0.0017$^{*}$   & 0.011 (0.072) & 0.0000$^{***}$ & 0.013 (0.103) & \\ \hline

\multicolumn{7}{l}{$^* p<0.05$; $^{**} p<0.01$; $^{***} p<0.001$} \\
\end{tabular}
\end{adjustbox}
\vspace*{-4pt}
\end{table}

\begin{table}[H]
\caption{Correlation of $TE_{SPO}$ with band power. $\widetilde{R}^2$ and $max(R^2)$ indicate the median and maximum of $R^2$ over recordings per condition, respectively.}
\label{tab:te_band_pow}
\begin{adjustbox}{center}
\begin{tabular}{lllcccccc}
\hline
\textbf{animal} & \textbf{level} & \textbf{direction} & \multicolumn{2}{c}{$\delta$} & \multicolumn{2}{c}{$\theta$} & \multicolumn{2}{c}{$\alpha$}\\
& & & $p$ & $\widetilde{R}^2$ ($max(R^2)$ & $p$ & $\widetilde{R}^2$ ($max(R^2)$ & $p$ & $\widetilde{R}^2$ ($max(R^2)$ \\ \hline
1 & iso 0.0 \% & PFC $\rightarrow$ V1 & 1.0000         & 0.024 (0.101) & 1.0000 & 0.034 (0.122) & 1.0000 & 0.061 (0.146) \\
  &            & V1 $\rightarrow$ PFC & 0.9662         & 0.030 (0.230) & 0.8389 & 0.009 (0.045) & 1.0000 & 0.008 (0.047) \\
  & iso 0.5 \% & PFC $\rightarrow$ V1 & 1.0000         & 0.033 (0.090) & 1.0000 & 0.073 (0.132) & 1.0000 & 0.013 (0.117) \\
  &            & V1 $\rightarrow$ PFC & 0.1319         & 0.001 (0.010) & 1.0000 & 0.018 (0.058) & 0.9097 & 0.001 (0.015) \\
  & iso 1.0 \% & PFC $\rightarrow$ V1 & 1.0000         & 0.009 (0.058) & 1.0000 & 0.009 (0.029) & 1.0000 & 0.012 (0.038) \\
  &            & V1 $\rightarrow$ PFC & 0.0947         & 0.002 (0.024) & 1.0000 & 0.003 (0.116) & 0.9996 & 0.004 (0.058) \\ \hline

2 & iso 0.0 \% & PFC $\rightarrow$ V1 & 1.0000         & 0.060 (0.102) & 1.0000 & 0.055 (0.121) & 1.0000 & 0.036 (0.052) \\
  &            & V1 $\rightarrow$ PFC & 0.9998         & 0.006 (0.038) & 1.0000 & 0.012 (0.056) & 0.9998 & 0.010 (0.040) \\
  & iso 0.5 \% & PFC $\rightarrow$ V1 & 1.0000         & 0.108 (0.157) & 1.0000 & 0.115 (0.166) & 1.0000 & 0.079 (0.160) \\
  &            & V1 $\rightarrow$ PFC & 1.0000         & 0.026 (0.149) & 1.0000 & 0.015 (0.052) & 0.6675 & 0.003 (0.009) \\
  & iso 1.0 \% & PFC $\rightarrow$ V1 & 0.0000$^{***}$ & 0.009 (0.095) & 1.0000 & 0.010 (0.037) & 1.0000 & 0.007 (0.030) \\
  &            & V1 $\rightarrow$ PFC & 0.0153         & 0.008 (0.023) & 1.0000 & 0.006 (0.035) & 0.7369 & 0.001 (0.017) \\

  &&&&&&&& \\
  &&&&&&&& \\ \hline
  \textbf{animal} & \textbf{level} & \textbf{site}  & \multicolumn{2}{c}{$\beta$}  & \multicolumn{2}{c}{$\gamma$} & \\
  &        &     & $p$ & $\widetilde{R}^2$ ($max(R^2$) & $p$ & $\widetilde{R}^2$ ($max(R^2)$ & \\ \hline
1 & iso 0.0 \% & PFC $\rightarrow$ V1 & 1.0000         & 0.069 (0.157) & 0.0352         & 0.005 (0.044) & \\
  &            & V1 $\rightarrow$ PFC & 0.9612         & 0.010 (0.040) & 0.0009$^{**}$  & 0.010 (0.028) & \\
  & iso 0.5 \% & PFC $\rightarrow$ V1 & 1.0000         & 0.033 (0.121) & 0.0611         & 0.031 (0.038) & \\
  &            & V1 $\rightarrow$ PFC & 0.9991         & 0.010 (0.039) & 0.0014$^{**}$  & 0.021 (0.028) & \\
  & iso 1.0 \% & PFC $\rightarrow$ V1 & 0.0000$^{***}$ & 0.004 (0.019) & 0.0000$^{***}$ & 0.007 (0.054) & \\
  &            & V1 $\rightarrow$ PFC & 0.0912         & 0.002 (0.063) & 0.0000$^{***}$ & 0.006 (0.109) & \\ \hline

2 & iso 0.0 \% & PFC $\rightarrow$ V1 & 1.0000         & 0.021 (0.040) & 0.3062         & 0.015 (0.036) & \\
  &            & V1 $\rightarrow$ PFC & 0.9998         & 0.002 (0.039) & 0.2046         & 0.002 (0.041) & \\
  & iso 0.5 \% & PFC $\rightarrow$ V1 & 1.0000         & 0.107 (0.132) & 0.6054         & 0.006 (0.013) & \\
  &            & V1 $\rightarrow$ PFC & 0.0186         & 0.004 (0.019) & 0.0000$^{***}$ & 0.022 (0.041) & \\
  & iso 1.0 \% & PFC $\rightarrow$ V1 & 0.9940         & 0.004 (0.039) & 0.0025$^{*}$   & 0.003 (0.027) & \\
  &            & V1 $\rightarrow$ PFC & 0.9122         & 0.002 (0.013) & 0.0000$^{***}$ & 0.002 (0.058) & \\ \hline

\multicolumn{7}{l}{$^* p<0.05$; $^{**} p<0.01$; $^{***} p<0.001$} \\
\end{tabular}
\end{adjustbox}
\vspace*{-4pt}
\end{table}

\paragraph*{S3 The effect of sampling on AIS-estimation}
\label{S3_ais_sampling}
{\bf AIS estimates for data sampled at different rates.}

\begin{figure}[H]
		\begin{adjustbox}{center}
	   		\includegraphics{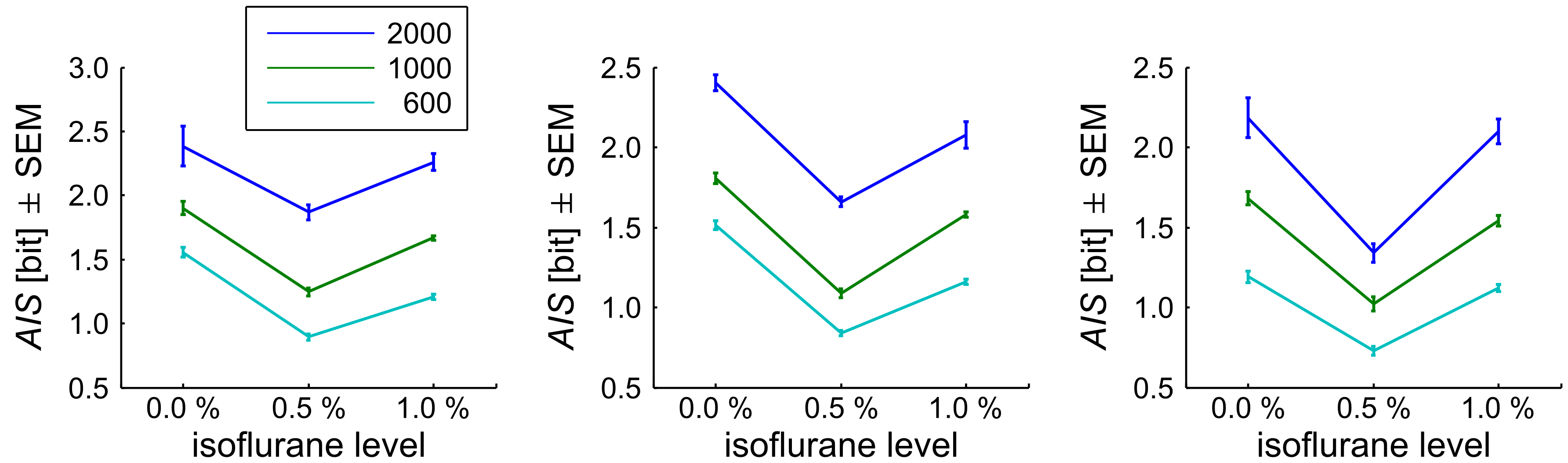}
		\end{adjustbox}
	\caption{{\bf Estimates of active information storage (AIS) from data sampled at different rates.}
    AIS estimates from three random recordings in animal 1 under three levels of Isoflurane; estimated from data with the sampling rate used for analysis in the present work (1000~Hz) and re-sampled at 2000 and 600~Hz respectively; note that qualitative results did not change due to re-sampling, but absolute estimates increased for higher sampling rates; the number of data points was held approximately constant by selecting a subset of trials for estimation such that the number of points entering the analysis was equal to the smallest number of points over all isoflurane levels.
    }
\label{fig:simulation_resampling}
\end{figure}

\paragraph*{S4 Theoretical and practical running time of nearest-neighbor based estimators} The KSG-estimator used in this work is computationally demanding, because as a nearest-neighbor based estimator it requires the execution of $k$-nearest-neighbor as well as range searches for all data points $N$. Hence, the algorithms used to perform these searches determine the asymptotic time complexity of the estimator. In this work, we used two different implementations published as part of the TRENTOOL toolbox: a CPU-based estimator for the estimation of $TE_{SPO}$ from epochs or trials of data \cite{lindner2011}, and a GPU-based estimator for the estimation of $TE_{SPO}$ from data pooled over epochs or trials \cite{wollstadt2014}. We used the CPU-based work-flow for the epoch-wise estimation of $TE_{SPO}$ and the GPU-based work-flow when testing for statistical significance of $TE_{SPO}$ to maximize the number of points entering the estimation. The CPU-based estimator uses a neighbor-search algorithm that---depending on data-properties---requires for neighbor-searches on all $N$ points at maximum $O(kN\log(N))$ time. The GPU-based estimator does not make use of fast data structures used for the CPU-based estimator, but uses a linear search which results in a worse time complexity of $O(dN^2)$. However, in the implementation used in this study \cite{wollstadt2014} the linear neighbor search is performed in parallel over points and problem instances, which significantly improves the overall running time when using the estimator on multiple problem instances---a comparison of the running times of the serial CPU-algorithm and the parallel GPU-algorithm can be found in \cite{wollstadt2014}.

Additionally, we used the KL-estimator for entropy estimation. This estimator requires the execution of a $k$-nearest-neighbor search for all data points $N$---we used the estimator's implementation published as part of the JIDT toolbox \cite{lizier2014JIDT}, which has a time complexity of $O(kN\log(N))$.

We also measured practical running times of the estimators to provide a point of reference when planning similar analyses. Approximate, average running times for the estimation of $AIS$, $TE_{SPO}$, and $H$ are presented in Table \ref{tab:running_time}. The presented running times include the estimation of each measure for both directions of interactions or recording site, for one recording; presented running times are averages over recordings. We measured the total time needed for estimation, including data preparation (e.g., the optimization of embedding parameters for the estimation of $TE_{SPO}$). All estimation procedures that did not require a GPU, were executed on a Intel(R) Xeon(R) CPU clocked at 2.90 GHz. The GPU-implementation of the $TE_{SPO}$ estimator was run on a Intel(R) Xeon(R) CPU clocked at 2.00 GHz and a NVIDIA GeForce GTX TITAN. Both machines were running 64-bit Ubuntu Linux. Note that the GPU estimation was performed on 50 trials only, because here the computational demand was higher due to more data points entering one estimation of $TE_{SPO}$ (pooled over epochs).

\begin{table}[H]
	\caption{Practical average running times for estimation of information-theoretic measures from one recording session and two recording sites or directions of interaction in animal 1.}
	\label{tab:running_time}
	\begin{tabular}{llr}
		\hline
		\textbf{measure} & \textbf{toolbox/implementation} & \textbf{mean running time [min]} \\ \hline
		$TE_{SPO}$  & TRENTOOL/GPU-implementation  & 2235.93 (314.27 SD) \\
		$TE_{SPO}$  & TRENTOOL/CPU-implementation  & 1871.51 (926.56 SD) \\
		$AIS$		& TRENTOOL/CPU-implementation  & 2.83 (1.51 SD)\\
		$H$			& JIDT/CPU-implementation	   & 1.55 (0.56 SD) \\ \hline
		\multicolumn{3}{l}{SD = standard deviation}
		\end{tabular}
		\vspace*{-4pt}
		\end{table}


\begin{thebibliography}{10}

\bibitem{dehaene2011GNW}
Dehaene S, Changeux JP.
\newblock {Experimental and theoretical approaches to conscious processing}.
\newblock Neuron. 2011;70(2):200--227.

\bibitem{tononi2004iit}
Tononi G.
\newblock An information integration theory of consciousness.
\newblock BMC Neurosci. 2004;5:42.

\bibitem{imas2005}
Imas OA, Ropella KM, Ward BD, Wood JD, Hudetz AG.
\newblock {Volatile anesthetics disrupt frontal-posterior recurrent information
  transfer at gamma frequencies in rat}.
\newblock Neurosci Lett. 2005;387(3):145--150.

\bibitem{hudetz2006}
Hudetz AG.
\newblock {Suppressing consciousness: mechanisms of general anesthesia}.
\newblock {Semin Anesth}. 2006;25(4):196--204.

\bibitem{alkire2008}
Alkire MT, Hudetz AG, Tononi G.
\newblock {Consciousness and anesthesia}.
\newblock Science. 2008;322(5903):876--880.

\bibitem{ku2011STE}
Ku SW, Lee U, Noh GJ, Jun IG, Mashour GA.
\newblock {Preferential inhibition of frontal-to-parietal feedback connectivity
  is a neurophysiologic correlate of general anesthesia in surgical patients}.
\newblock PLoS One. 2011;6(10):e25155.

\bibitem{lee2013}
Lee U, Ku S, Noh G, Baek S, Choi B, Mashour GA.
\newblock {Disruption of frontal-parietal communication by ketamine, propofol,
  and sevoflurane.}
\newblock Anesthesiology. 2013;118(6):1264--1275.

\bibitem{jordan2013}
Jordan D, Ilg R, Riedl V, Schorer A, Grimberg S, Neufang S, et~al.
\newblock {Simultaneous electroencephalographic and functional magnetic
  resonance imaging indicate impaired cortical top-down processing in
  association with anesthetic-induced unconsciousness.}
\newblock Anesthesiology. 2013;119(5):1031--1042.

\bibitem{untergehrer2014}
Untergehrer G, Jordan D, Kochs EF, Ilg R, Schneider G.
\newblock {Fronto-parietal connectivity is a non-static phenomenon with
  characteristic changes during unconsciousness}.
\newblock PLoS One. 2014;9(1):e87498.

\bibitem{schreiber2000}
Schreiber T.
\newblock {Measuring information transfer}.
\newblock Phys Rev Lett. 2000;85(2):461--464.

\bibitem{Ay2008}
Ay N, Polani D.
\newblock {Information flows in causal networks}.
\newblock Adv Complex Syst. 2008;11:17--41.

\bibitem{lizier2010}
Lizier JT, Prokopenko M.
\newblock {Differentiating information transfer and causal effect}.
\newblock Eur Phys J B. 2010;73:605--615.

\bibitem{chicharro2012}
Chicharro D, Ledberg A.
\newblock When two become one: the limits of causality analysis of brain
  dynamics.
\newblock PLoS One. 2012;7:e32466.

\bibitem{krasowski1999}
Krasowski M, Harrison N.
\newblock {General anaesthetic actions on ligand-gated ion channels}.
\newblock Cell Mol Life Sci. 1999;55(10):1278--1303.

\bibitem{wibral2013timing}
Wibral M, Pampu N, Priesemann V, Siebenh{\"u}hner F, Seiwert H, Lindner M,
  et~al.
\newblock {Measuring information-transfer delays}.
\newblock PLoS One. 2013;8(2):e55809.

\bibitem{wollstadt2015IEEE}
Wollstadt P, Sellers KK, Hutt A, Fr{\"o}hlich F, Wibral M.
\newblock Anesthesia-related changes in information transfer may be caused by
  reduction in local information generation.
\newblock In: Engineering in Medicine and Biology Society (EMBC), 2015 37th
  Annual International Conference of the IEEE. IEEE; 2015. p. 4045--4048.

\bibitem{nsb}
Nemenman I, Shafee F, Bialek W.
\newblock {Entropy and Inference, Revisited}.
\newblock In: {Advances in Neural Information Processing Systems 14}.
  Cambridge, MA: MIT Press; 2002. p. 471--478.

\bibitem{nsb2}
Nemenman I, Shafee F, van Steveninck RR.
\newblock {Entropy and information in neural spike trains: progress on the
  sampling problem}.
\newblock Phys Rev E. 2004;69:056111.

\bibitem{lizier2012LAIS}
Lizier JT, Prokopenko M, Zomaya AY.
\newblock {Local measures of information storage in complex distributed
  computation}.
\newblock Inform Sciences. 2012;208:39--54.

\bibitem{archer2014}
Archer E, Park IM, Pillow JW.
\newblock {Bayesian entropy estimation for countable discrete distributions}.
\newblock J Mach Learn Res. 2014;15:2833--2868.

\bibitem{ragwitz2002}
Ragwitz M, Kantz H.
\newblock {Markov models from data by simple nonlinear time series predictors
  in delay embedding spaces.}
\newblock Phys Rev E. 2002;65(5):056201.

\bibitem{vicente2011TE}
Vicente R, Wibral M, Lindner M, Pipa G.
\newblock {Transfer entropy--a model-free measure of effective connectivity for
  the neurosciences}.
\newblock J Comput Neurosci. 2011;30(1):45--67.

\bibitem{lindner2011}
Lindner M, Vicente R, Priesemann V, Wibral M.
\newblock {{{TRENTOOL}}: a Matlab open source toolbox to analyse information
  flow in time series data with transfer entropy.}
\newblock BMC Neurosci. 2011;12(1):119.

\bibitem{barnett2011}
Barnett L, Seth AK.
\newblock {Behaviour of Granger causality under filtering: Theoretical
  invariance and practical application}.
\newblock J Neurosci Methods. 2011;201(2):404--419.


\bibitem{florin2010}
Florin E, Gross J, Pfeifer J, Fink GR, Timmermann L.
\newblock {The effect of filtering on Granger causality based multivariate
  causality measures}.
\newblock Neuroimage. 2010;50(2):577--588.


\bibitem{oostenveld2011FT}
Oostenveld R, Fries P, Maris E, Schoffelen JM.
\newblock FieldTrip: open source software for advanced analysis of {{MEG}},
  {{EEG}}, and invasive electrophysiological data.
\newblock Comput Intell Neurosci. 2011;2011:1--9.

\bibitem{gomez2014}
G{\'o}mez C, Lizier JT, Schaum M, Wollstadt P, Gr{\"u}tzner C, Uhlhaas PJ,
  et~al.
\newblock {Reduced predictable information in brain signals in autism spectrum
  disorder}.
\newblock Frontiers in Neuroinformatics. 2014;8:9.

\bibitem{brodski2016}
Brodski-Guerniero A, Paasch GF, Wollstadt P, Oezdemir I, Lizier JT, Wibral M.
\newblock Activating task relevant prior knowledge increases active information
  storage in content specific brain areas;
\newblock 2016. Preprint. Available from: bioRxiv: \url{https://doi.org/10.1101/089300}. Cited 24 March 2017.

\bibitem{wibral2015a}
Wibral M, Lizier JT, Priesemann V.
\newblock {Bits from brains for biologically inspired computing}.
\newblock Frontiers in Robotics and AI. 2015;2:5.


\bibitem{purdon2013}
Purdon PL, Pierce ET, Mukamel EA, Prerau MJ, Walsh JL, Wong KFK, et~al.
\newblock {Electroencephalogram signatures of loss and recovery of
  consciousness from propofol}.
\newblock PNAS. 2013;110(12):E1142--E1151.

\bibitem{felleman1991}
Felleman DJ, {Van Essen} DC.
\newblock {Distributed hierarchical processing in the primate cerebral cortex}.
\newblock Cereb Cort. 1991;1(1):1--47.

\bibitem{salin1995}
Salin PA, Bullier J.
\newblock {Corticocortical connections in the visual system: structure and
  function}.
\newblock Physiol Rev. 1995;75(1):107--154.

\bibitem{tomioka2005}
Tomioka R, Okamoto K, Furuta T, Fujiyama F, Iwasato T, Yanagawa Y, et~al.
\newblock {Demonstration of long-range GABAergic connections distributed
  throughout the mouse neocortex}.
\newblock Eur J Neurosci. 2005;21(6):1587--1600.


\bibitem{clark2013}
Clark A.
\newblock {Whatever next? Predictive brains, situated agents, and the future of
  cognitive science}.
\newblock Behav Brain Sci. 2013;36(03):181--204.

\bibitem{hohwy2013predictivemind}
Hohwy J.
\newblock {The Predictive Mind}.
\newblock New York: Oxford University Press; 2013.

\bibitem{hawkins2007}
Hawkins J, Blakeslee S.
\newblock {On Intelligence}.
\newblock New York: Macmillan; 2007.

\bibitem{bastos2012}
Bastos AM, Usrey WM, Adams RA, Mangun GR, Fries P, Friston KJ.
\newblock {Canonical microcircuits for predictive coding}.
\newblock Neuron. 2012;76(4):695--711.

\bibitem{mashour2014}
Mashour GA.
\newblock {Top-down mechanisms of anesthetic-induced unconsciousness}.
\newblock Front Syst Neurosci. 2014;8:115.

\bibitem{brodski2015}
Brodski A, Paasch GF, Helbling S, Wibral M.
\newblock {The faces of predictive coding}.
\newblock J Neurosci. 2015;35(24):8997--9006.

\bibitem{rivolta-ketamine-2015}
Rivolta D, Heidegger T, Scheller B, Sauer A, Schaum M, Birkner K, et~al.
\newblock Ketamine dysregulates the amplitude and connectivity of
  high-frequency oscillations in cortical--subcortical networks in humans:
  evidence from resting-state magnetoencephalography-recordings.
\newblock Schizophr Bull. 2015;41(5):1105--1114.

\bibitem{spinney2016}
Spinney RE, Prokopenko M, Lizier JT.
\newblock {Transfer entropy in continuous time, with applications to jump and
  neural spiking processes};
\newblock 2016. Preprint. Available from: arXiv:161008192v1. Cited 27 December
  2016.

\bibitem{frohlich2010endogenous}
Fr{\"o}hlich F, McCormick DA.
\newblock Endogenous electric fields may guide neocortical network activity.
\newblock Neuron. 2010;67(1):129--143.

\bibitem{tao2005szemer}
Tao T.
\newblock Szemer\'edi's regularity lemma revisited;
\newblock 2005. Preprint. Available from: arXiv:math/0504472. Cited 24 November
  2016.

\bibitem{roux2012}
Roux F, Wibral M, Mohr HM, Singer W, Uhlhaas PJ.
\newblock {Gamma-band activity in human prefrontal cortex codes for the number
  of relevant items maintained in working memory}.
\newblock Journal of Neuroscience. 2012;32(36):12411--12420.

\bibitem{raw_data}
Sellers KK, Fr{\"o}hlich F. Data from: Breakdown of local information
  processing may underlie isoflurane anesthesia effects; 2017.
\newblock Dryad Digital Repository. Available from: \url{http://dx.doi.org/10.5061/dryad.kk40s} (available after acceptance of the manuscript).

\bibitem{sellers2013}
Sellers KK, Bennett DV, Hutt A, Fr{\"o}hlich F.
\newblock {Anesthesia differentially modulates spontaneous network dynamics by
  cortical area and layer}.
\newblock J Neurophysiol. 2013;110(12):2739--2751.

\bibitem{sellers2015}
Sellers KK, Bennett DV, Hutt A, Williams J, Fr{\"o}hlich F.
\newblock {Awake versus anesthetized: layer-specific sensory processing in
  visual cortex and functional connectivity between cortical areas}.
\newblock J Clin Neurophysiol. 2015;113:3798--3815.

\bibitem{cover2006}
Cover TM, Thomas JA.
\newblock {Elements of Information Theory}.
\newblock New York: Wiley; 2006.

\bibitem{wibral2014springer}
Wibral M, Vicente R, Lindner M.
\newblock {Transfer entropy in neuroscience}.
\newblock In: Wibral M, Vicente R, Lizier JT editors. {Directed Information Measures in Neuroscience}. Heidelberg:
  Springer; 2014. pp. 3--58.

\bibitem{WilliamsII}
Williams PL, Beer RD.
\newblock {Generalized measures of information transfer};
\newblock 2011. Preprint. Available from: arXiv:11021507. Cited 24 November
  2016.

\bibitem{takens1981}
Takens F.
\newblock Detecting strange attractors in turbulence.
\newblock In: {Dynamical Systems and Turbulence, Warwick 1980}. vol. 898 of
  {Lecture Notes in Mathematics}. Berlin: Springer; 1981. p. 366--381.

\bibitem{wibral2014LAIS}
Wibral M, Lizier JT, V{\"o}gler S, Priesemann V, Galuske R.
\newblock {Local active information storage as a tool to understand distributed
  neural information processing}.
\newblock Front Neuroinform. 2014;8.

\bibitem{lizier2014JIDT}
Lizier JT.
\newblock {{{JIDT}}: an information-theoretic toolkit for studying the dynamics
  of complex systems}.
\newblock Front Robot AI. 2014;1(11).

\bibitem{faes2014springer}
Faes L, Porta A.
\newblock {Conditional entropy-based evaluation of information dynamics in
  physiological systems}.
\newblock In: Wibral M, Vicente R, Lizier JT editors. {Directed Information Measures in Neuroscience}. Heidelberg:
  Springer; 2014. pp. 61--86.

\bibitem{staniek2008STE}
Staniek M, Lehnertz K.
\newblock {Symbolic transfer entropy}.
\newblock Phys Rev Lett. 2008;100:158101.

\bibitem{pompe2011MIT}
Pompe B, Runge J.
\newblock {Momentary information transfer as a coupling measure of time
  series}.
\newblock Phys Rev E. 2011;83(5):051122.

\bibitem{kraskov2004}
Kraskov A, St{\"o}gbauer H, Grassberger P.
\newblock {Estimating mutual information.}
\newblock Phys Rev E. 2004;69(6):066138.

\bibitem{frenzel2007}
Frenzel S, Pompe B.
\newblock {Partial mutual information for coupling analysis of multivariate
  time series.}
\newblock {Phys Rev Lett}. 2007;99(20):204101.

\bibitem{kraskov2004thesis}
Kraskov A.
\newblock Synchronization and interdependence measures and their application to
  the electroencephalogram of epilepsy patients and clustering of data.
\newblock PhD Thesis. University of Wuppertal; 2004.

\bibitem{khan2007MIest}
Khan S, Bandyopadhyay S, Ganguly AR, Saigal S, {Erickson III} DJ, Protopopescu
  V, et~al.
\newblock {Relative performance of mutual information estimation methods for
  quantifying the dependence among short and noisy data}.
\newblock Phys Rev E. 2007;76(2):026209.

\bibitem{wollstadt2014}
Wollstadt P, Mart{\'i}nez-Zarzuela M, Vicente R, D{\'i}az-Pernas FJ, Wibral M.
\newblock {Efficient transfer entropy analysis of non-stationary neural time
  series}.
\newblock PLoS One. 2014;9(7):e102833.

\bibitem{kozachenko1987}
Kozachenko LF, Leonenko NN.
\newblock {On statistical estimation of entropy of random vector}.
\newblock Problems Inform Transm. 1987;23:95--101.

\bibitem{wolpert1995estimating}
Wolpert DH, Wolf DR.
\newblock {Estimating functions of probability distributions from a finite set
  of samples}.
\newblock Phys Rev E. 1995;52(6):6841.

\bibitem{faes2011nonuniform}
Faes L, Nollo G, Porta A.
\newblock {Information-based detection of nonlinear Granger causality in
  multivariate processes via a nonuniform embedding technique}.
\newblock Phys Rev E. 2011;83(5):051112.

\bibitem{anderson2003pANOVA}
Anderson M, Braak CT.
\newblock {Permutation tests for multi-factorial analysis of variance}.
\newblock J Stat Comput Simul. 2003;73(2):85--113.

\bibitem{suckling2004pANOVA}
Suckling J, Bullmore E.
\newblock {Permutation tests for factorially designed neuroimaging
  experiments}.
\newblock Hum Brain Mapp. 2004;22(3):193--205.

\bibitem{helbling2015}
Helbling S.
\newblock {Advances in MEG methods and their applications to investigate
  auditory perception}.
\newblock PhD Thesis. Goethe-University, Frankfurt; 2015.

\bibitem{helbling2017github}
Helbling S. {permANOVA toolbox}; 2017.
\newblock Available from: https://github.com/sashel/permANOVA. Cited 26 January 2017.

\bibitem{aarts2014}
Aarts E, Verhage M, Veenvliet JV, Dolan CV, van~der Sluis S.
\newblock {A solution to dependency: using multilevel analysis to accommodate
  nested data}.
\newblock Nat Neurosci. 2014;17(4):491--496.

\bibitem{fahrmeir2007regression}
Fahrmeir L, Kneib T, Lang S.
\newblock {Regression}.
\newblock Berlin: Springer; 2007.

\bibitem{Rlanguage}
{R Development Core Team}. {R: a language and environment for statistical
  computing}.
\newblock R Foundation for Statistical Computing; 2008.
\newblock Vienna, Austria.
\newblock http://www.R-project.org.

\bibitem{lme4}
Bates D, M{\"a}chler M, Bolker B, Walker S.
\newblock {Fitting linear mixed-effects models using {lme4}}.
\newblock J Stat Softw. 2015;67(1):1--48.

\bibitem{barr2013}
Barr DJ, Levy R, Scheepers C, Tily HJ.
\newblock {Random effects structure for confirmatory hypothesis testing: keep
  it maximal}.
\newblock J Mem Lang. 2013;68(3):255--278.

\bibitem{vakorin2014springer}
Vakorin VA, Krakovska O, McIntosh AR
\newblock {On complexity and phase effects in reconstructing the directionality
of coupling in non-linear systems}.
\newblock In: Wibral M, Vicente R, Lizier JT editors. {Directed Information Measures in Neuroscience}. Heidelberg:
  Springer; 2014. pp. 137--158.

\end{thebibliography}
\end{document}